\PassOptionsToPackage{sort&compress}{natbib}
\documentclass[numbers]{assets/matterlab}

\usepackage{microtype}
\usepackage{graphicx}
\usepackage{booktabs}
\usepackage{float}
\usepackage{xurl}
\usepackage{hyperref}
\usepackage{titlesec}

\usepackage{amsmath}
\usepackage{amssymb}
\usepackage{mathtools}
\usepackage{amsthm}
\usepackage{bm}

\usepackage[noabbrev,nameinlink]{cleveref}

\usepackage{amsmath,amsfonts,bm}

\def\eqref#1{equation~\ref{#1}}

\def\1{\bm{1}}

\DeclareMathAlphabet{\mathsfit}{\encodingdefault}{\sfdefault}{m}{sl}
\SetMathAlphabet{\mathsfit}{bold}{\encodingdefault}{\sfdefault}{bx}{n}

\usepackage{orcidlink}
\usepackage{comment}
\usepackage[version=4]{mhchem}
\usepackage{longtable}
\usepackage{array}
\renewcommand{\arraystretch}{1.7}
\usepackage{multirow}           
\usepackage{makecell}           
\usepackage{amsfonts}           
\usepackage{nicefrac}
\usepackage{duckuments}         
\usepackage{thmtools,thm-restate}
\usepackage{enumitem}
\usepackage{xcolor,colortbl}
\usepackage{tikz}
\usepackage{tikz-cd}
\usepackage{caption}
\usepackage{subcaption}
\usepackage{listings}
\usepackage{gensymb}
\usepackage{textcomp} 
\usepackage[inkscapelatex=false]{svg}

\usepackage[type1]{cinzel}

\newcommand{\addressCHEM}{Department of Chemistry, University of Toronto,  80 St. George St., Toronto, ON M5S 3H6, Canada}
\newcommand{\addressAC}{Acceleration Consortium, 700 University Ave., Toronto, ON M7A 2S4, Canada}
\newcommand{\addressCS}{Department of Computer Science, University of Toronto, 40 St George St., Toronto, ON M5S 2E4, Canada}
\newcommand{\addressVECTOR}{Vector Institute for Artificial Intelligence, W1140-108 College St., Schwartz Reisman Innovation Campus, Toronto, ON M5G 0C6, Canada}
\newcommand{\addressMSE}{Department of Materials Science \& Engineering, University of Toronto, 184 College St., Toronto, ON M5S 3E4, Canada}
\newcommand{\addressCHEMENG}{Department of Chemical Engineering \& Applied Chemistry, University of Toronto, 200 College St., Toronto, ON M5S 3E5, Canada}
\newcommand{\addressMEDICALSCI}{Institute of Medical Science, 1 King's College Circle, Medical Sciences Building, Room 2374, Toronto, ON M5S 1A8, Canada}
\newcommand{\addressCIFAR}{Canadian Institute for Advanced Research (CIFAR), 661 University Ave., Toronto,
ON M5G 1M1, Canada}
\newcommand{\addressNVIDIA}{NVIDIA, 431 King St W \#6th, Toronto, ON M5V 1K4, Canada}

\newcommand{\acknowAC}{This research is part of the University of Toronto’s Acceleration Consortium, which receives funding from the CFREF-2022-00042 Canada First Research Excellence Fund. }

\newcommand{\acknowDARPA}{This work was supported by the Defense Advanced Research Projects Agency (DARPA) under Agreement No.~HR0011262E022.}

\providecommand{\doi}[1]{\href{https://doi.org/#1}{doi:#1}}

\setcitestyle{square,comma}
\usepackage{flafter}   
\newcommand{\bench}{{\cinzel MolDraft}}

\newcommand{\citeDataCODPrimary}{\citep{husebye2001,akiba2001,yurkerwich2008,dohi2010,rong2014,prasad2015,hill2018,hartmann2019,hokamp2019,kadu2019,hartmann2021,luo2023,gon2024,baryshnikova2026}}
\newcommand{\citeDataCSDPrimary}{\citep{hering2015,gao2021,disanto2018,jones1998,shimada1996,huckstadt1997,cybulski2017,marchetti2008,lu2013,nikiforov2008,perez2021,manakin2018,manakin2023,malassa2010,rheingold2019,schendzielorz2016,wieder2011,askevold2010,lake2016,beck2001,sukhikh2020,sangtrirutnugul2006,noschangcabral2023,huang2022,schaub2006,west2009,dejesussilva2021,bierschenk1987,abram1999,schultz2011,rosset1990,gupta2022,yu2022,kakaliou1999,ortner2000,krause1998,li2008,davies2000,husebo1995,reckziegel2020,minnick2023}}

\usepackage{soul}  
\newcommand{\Note}[1]{}
\renewcommand{\Note}[1]{#1}  

\title{Can Language Models Build Molecules in Three Dimensions?}

\author[1]{Hanyong Xu}
\author[2,7,9,\dagger]{Yunheng Zou}
\author[10,\dagger]{Olalla Nieto Faza}
\author[1,7]{Juan Bernardo Pérez Sánchez}
\author[1]{Aiwei Yin}
\author[1,2,3,4,5,6,7,8,9,*]{Al\'an Aspuru-Guzik}
\author[1,2,6,*]{Varinia Bernales}

\affiliation[1]{\addressCHEM}
\affiliation[2]{\addressCS}
\affiliation[3]{\addressMSE}
\affiliation[4]{\addressCHEMENG}
\affiliation[5]{\addressMEDICALSCI}
\affiliation[6]{\addressAC}
\affiliation[7]{\addressVECTOR}
\affiliation[8]{\addressCIFAR}
\affiliation[9]{\addressNVIDIA}
\affiliation[10]{Department of Organic Chemistry, Universidade de Vigo, Spain}

\contribution[\dagger]{These authors contributed equally}

\abstract{
Chemical drawings convey which atoms are bonded and how stereocentres are configured, but not molecular shape. Every simulation starts from three-dimensional coordinates, and building them from a drawing remains a manual step. We introduce \bench{}, a curated benchmark of 300 drawing–structure pairs spanning natural products, hypervalent compounds, and coordination complexes. Given an image and a short instruction, an AI system must generate a complete set of atomic coordinates, which are checked for molecular identity, stereochemical configuration, and coarse ligand geometry. We evaluate nine multimodal language models using the same agent setup. The share of molecules built correctly, the resolve rate, ranges from 26.3\% to 85.0\% and falls as molecules grow, most sharply above about 80 atoms. Only 30 of the 300 molecules are resolved by all nine models, and literature structures, mostly metal complexes, are the hardest source for every model. Language models can thus build many molecules in three dimensions, but not yet reliably reproduce the coordination chemistry of the research literature.

}

\date{\today}
\correspondence{\email{varinia@bernales.org} and \email{alan@aspuru.com}}

\begin{document}

\enlargethispage{2\baselineskip}
\maketitle
\clearpage


\section{Introduction}
\label{sec:intro}

Two-dimensional molecular drawings are central to chemical communication because they convey molecular identity, functional groups, and stereochemical relationships in a compact, standardized form suitable for publication, comparison, and reaction schemes~\citep{evans2014chemdraw,zhang2026molecular}. By emphasizing connectivity and bond changes without specifying a particular conformation, they support synthesis planning and mechanistic discussion. In contrast, three-dimensional geometries reveal molecular shape, steric interactions, and coordination environments, and provide the inputs for docking, geometry optimization, and quantum-chemical calculations~\citep{trott2010vina,frisch2016gaussian16,neese2025orca}. Consequently, chemists often need to translate published drawings into three-dimensional models, a process that requires manual atom placement, model construction, and repeated correction.

Reliable automated 2D-to-3D conversion could bridge these representations. It would reduce this burden and connect chemical drawings directly to computational workflows while preserving the connectivity, stereochemistry, and ligand arrangement of the intended species. Such conversion involves two successive steps, image-to-graph recognition and conformer generation~\citep{qian2023molscribe,riniker2015etkdg}. Despite progress in these areas, reliable reconstruction across diverse chemical classes remains difficult: systems must interpret abbreviated functional groups, stereochemical and perspective cues, and metal--ligand bonding before constructing an appropriate three-dimensional geometry. Even when molecular connectivity is correctly identified, embedding algorithms can struggle with geometrically constrained systems, such as strained rings~\citep{qian2023molscribe,demuth2026strained}. These pipelines are also fixed sequences. When an error in recognition passes silently into embedding, no step checks whether the final structure is the species the drawing intends.

Language-model agents offer a different route, because they can interpret a request, plan a workflow, and inspect and revise their own results~\citep{bran2024chemcrow,boiko2023coscientist}. They have moved quickly from answering chemistry questions to acting on them. They now plan, execute, and analyse computational workflows from a natural-language request: they call cheminformatics libraries, write input files, and launch electronic-structure calculations. Yet in most such agents the three-dimensional structure is still built through the same pipeline: the model writes a SMILES string, or retrieves one, and a toolkit such as RDKit~\citep{rdkit} or Open Babel~\citep{oboyle2011openbabel} embeds it. This route works well for organic molecules, for which the line notation was designed, but it inherits the notation's limits. RDKit supports dative bonds and explicit square-planar stereochemistry in SMILES~\citep{rdkit}. Multicentre and $\eta$-coordination can still be difficult to translate into coordinates, and omitting or mistranscribing a stereodescriptor can lose a configuration fixed by a wedge in the drawing. {\cinzel El Agente Q}~\citep{zou2025elagente}, a hierarchical multi-agent system for quantum chemistry, illustrates both the promise and the weak point of this design. Across university-level exercises, it completed more than 87\% of tasks on average, yet its inorganic exercise scored lower when the molecules were supplied as SMILES ($78 \pm 9\%$) than when they were supplied as coordinates ($84 \pm 7\%$). The authors attributed this difference to SMILES-to-coordinate conversion being less reliable for complex coordination environments, non-standard stereochemistry, and varied oxidation states.

{\cinzel El Agente Estructural}~\citep{choi2026estructural} changed the division of labour by placing structure construction under the direct control of a vision--language model. The agent combines geometry-aware tools for substitution, ligand exchange, and stereochemically controlled editing with rendered views of the structure it builds, and it showed chemically meaningful construction of organometallic stereoisomers, reaction intermediates, and transition-state guesses across a series of case studies. The same work identified the remaining bottleneck in the model, not the tools. When asked only to name the atoms bonded to a target atom from rendered views, five frontier vision--language models lost accuracy steadily with molecular size, and all struggled above about 60 atoms. Models also often failed to recognize motifs as simple as a six-membered ring in two-dimensional reaction diagrams, and the authors concluded that perception and chemical reasoning remain the limiting steps. Related studies point in the same way: frontier models without fine-tuning generate poor molecular geometries~\citep{zhang2026molecular,cavanagh2026geometry}, and spatial reasoning degrades as structural operations grow more complex~\cite{lv2025atomworld}. The open question is therefore no longer whether an agent can operate the tools, but whether the underlying model can read a chemical drawing and build the molecule it depicts correctly, at the scale of research structures. That capability has not been measured systematically across organic, main-group, and coordination chemistry.

\begin{figure}[H]
    \centering
    \captionsetup{skip=4pt}
    \includegraphics[width=0.99\textwidth]{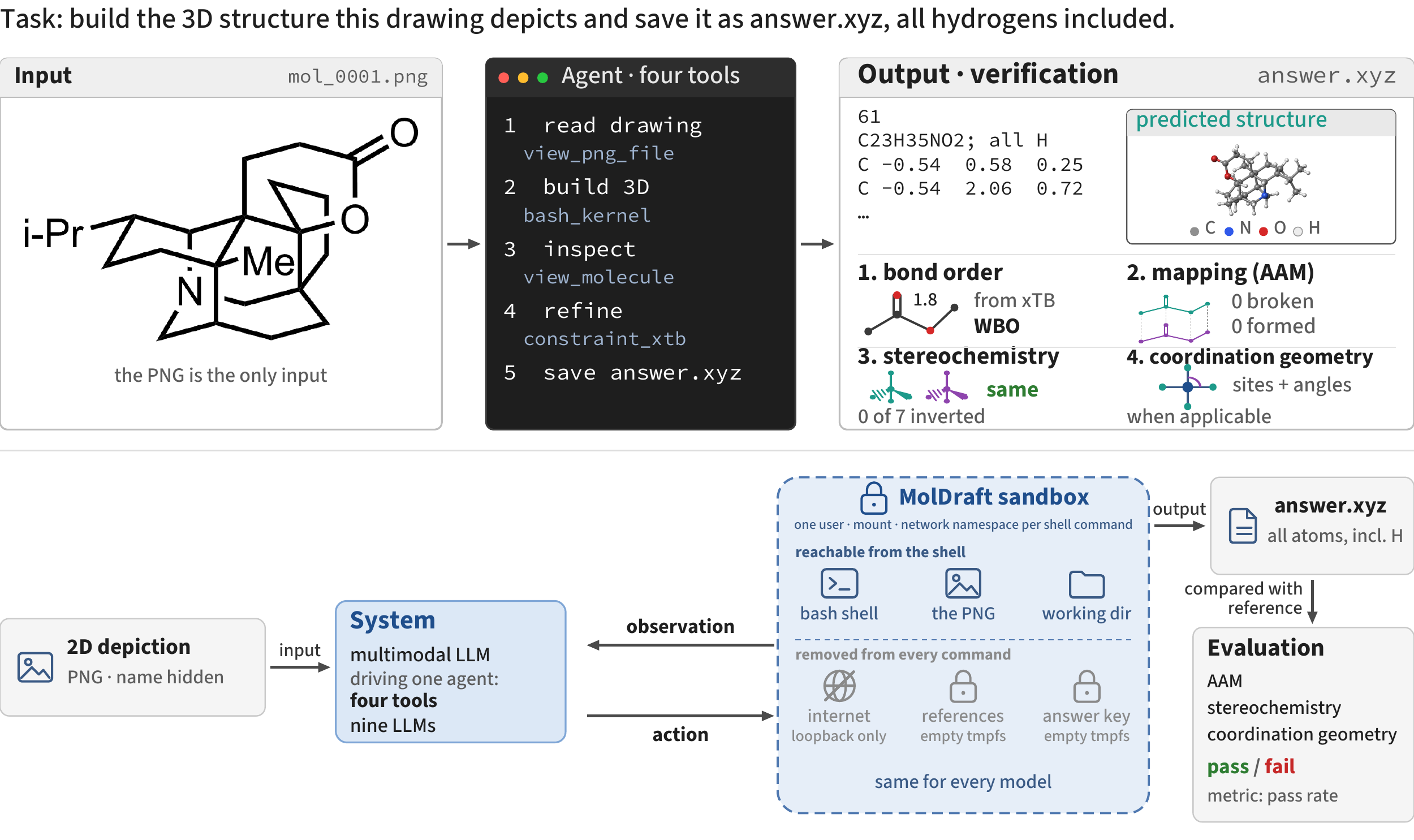}
    \caption{\textbf{\bench{} task and evaluation pipeline.}
    \emph{Top:} an anonymised drawing of daphnilactone A (left) is the only input. The agent reads it, builds a three-dimensional structure with four tools for image viewing, shell execution, structure inspection, and xTB relaxation, and saves Cartesian coordinates for every atom, hydrogens included (centre). An independent verifier checks the saved structure (right).
    \emph{Bottom:} the evaluation loop. A multimodal language model drives the agent inside a sandbox without internet or reference access, and the saved structure is compared with the reference. The verifier checks composition, connectivity, stereochemistry, and coarse ligand geometry (\Cref{sec:agentsetup,sec:task}).}
    \label{fig:overview}
\end{figure}

To assess how reliably systems perform this conversion, we introduce \bench{}, a benchmark for reconstructing three-dimensional molecular structures from drawings across organic, hypervalent, and coordination chemistry (\Cref{fig:overview}).
Given an anonymised image of a drawing, a system must return Cartesian coordinates for every atom, including hydrogens.
The task tests whether a system preserves molecular identity and the specified stereochemistry during conversion. It applies to conventional pipelines, specialised models, and language-model agents.

All 300 input drawings and reference three-dimensional structures were manually reviewed for molecular identity and depicted stereochemistry (\Cref{app:validation}).
For the agent evaluation, each model receives the same four tools and task instructions and operates in a sandbox without internet or reference access.
An independent verifier checks the submitted structure's elemental composition, connectivity, stereochemistry, and coarse ligand geometry (\Cref{sec:task}).
Because the verifier is separate from the construction, models can be compared consistently on a demanding chemical task whose outcomes remain inexpensive to check~\citep{jimenez2024swebench}.
Each molecule is annotated by geometry, size, and complexity, so the results show which molecular features remain challenging.

\begingroup
\emergencystretch=3em
We evaluate nine frontier multimodal models: \texttt{gpt-6-astra}~\citep{openai2026astra}, \texttt{gpt-6.1-sol}~\citep{openai2026sol61}, \texttt{gpt-6-sol}~\citep{openai2026sol}, \texttt{gpt-6-luna}~\citep{openai2026luna}, \texttt{claude-opus-5}~\citep{anthropic2026opus5}, \texttt{claude-opus-5-5}~\citep{anthropic2026opus55}, \texttt{claude-fable-5-1}~\citep{anthropic2026fable51}, \texttt{qwen3.8-max}~\citep{qwen2026max38}, and \texttt{kimi-k3}~\citep{moonshot2026kimi3}. Resolve rates range from 26.3\% to 85.0\% over all 300 molecules, and only 30 of the 300 molecules are resolved by all nine models (\Cref{tab:results}).
All models perform worse on metal-containing molecules, with an average gap of 11.2 percentage points. Structures directly taken from the literature are the lowest-scoring source group for all nine models.
The benchmark, submitted structures, and verification results are publicly available at \url{https://github.com/the-matter-lab/MolDraft}; the archived agent session logs are linked in \Cref{app:availability}.
\par
\endgroup

\section{\bench{}}
\label{sec:bench}


\bench{} pairs two-dimensional molecular drawings with reference three-dimensional structures obtained from public databases and the chemical literature. The task is to generate complete Cartesian coordinates from each drawing. An independent verifier then checks the structure's elemental composition, connectivity, and stereochemical configuration.

\subsection{Data selection}
\label{sec:construction}

\begin{figure}[H]
    \centering
    \captionsetup{skip=4pt}
    \includegraphics[width=0.72\textwidth]{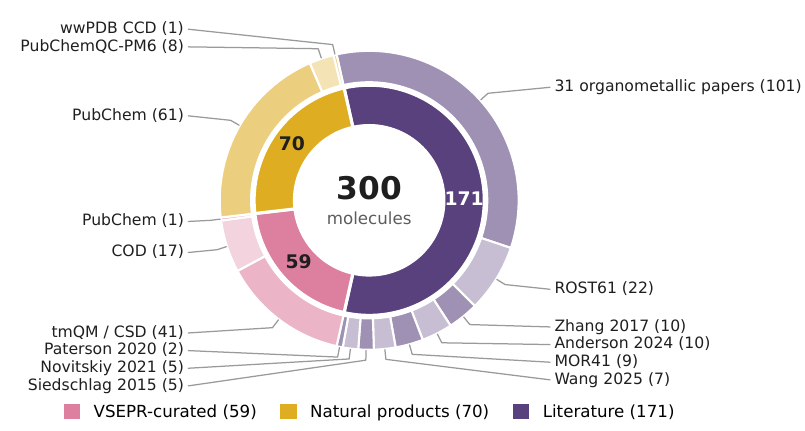}
    \caption{\textbf{The 300 molecules of \bench{} by source.} \emph{Inner ring:} the three sources of structures. \emph{Outer ring:} where their reference coordinates come from. For the VSEPR-curated entries and the natural products this is the database. For the 171 literature entries it is the paper or benchmark set, and the 31 papers that supply the organometallic figures are drawn as a single segment. Counts are in parentheses. References for every source are in \Cref{tab:sources,tab:papers}.}
    \label{fig:sources}
\end{figure}

\textbf{Coverage.}
The 300 molecules in \bench{} were selected to probe complementary demands: local geometric interpretation, complex organic connectivity and stereochemistry, and challenging coordination environments encountered in research papers.
Together they include 52 elements, 15 local geometry classes described using the valence-shell electron-pair repulsion (VSEPR) framework~\citep{gillespie1992vsepr}, coordination numbers from 2 to 8, and sizes from 8 to 106 heavy atoms. Of the 300 molecules, 201 contain a metal.
The three sources in \Cref{fig:sources} therefore broaden the evaluation beyond routine organic structures.

\textbf{VSEPR-curated structures (59).}
Here a class is a local arrangement of coordination sites around a central atom, such as bent, linear, tetrahedral, square planar, or octahedral.
Candidates were collected separately for each class so that common organic geometries would not dominate the set or displace less common hypervalent and metal coordination environments.
VSEPR-style labels provide an interpretable taxonomy of these arrangements and expose contrasts such as tetrahedral versus square-planar coordination. They are not used as a universal electronic-structure model for transition-metal bonding.
Reference coordinates come from PubChem, the Crystallography Open Database (COD), and tmQM, whose Cambridge Structural Database (CSD) structures were optimised using GFN2-xTB~\citep{kim2025pubchem,grazulis2012cod,balcells2020tmqm,groom2016csd}.
An automated script converted the reference XYZ coordinates into CDXML drawings, which were rendered in ChemDraw~\citep{evans2014chemdraw} and checked against the references (\Cref{app:depictions,app:validation}).
Within the available classes, selection favoured \textbf{demanding representatives}, including hypervalent centres and complexes with several ligand sites, while retaining coverage of every class (\Cref{app:construction}).

\textbf{Natural products (70).}
Targets of total syntheses were selected from SynArchive~\citep{synarchive} for their structural complexity, including polycyclic frameworks and multiple stereocentres.
The inputs are SynArchive's drawings, paired with PubChem or PubChemQC-PM6 reference conformers~\citep{kim2025pubchem,nakata2020}, except for chlorophyll~$a$, whose reference is wwPDB component CLA.
They test whether a system can construct complex organic molecules, so that the benchmark's difficulty is not defined solely by metal content.

\textbf{Structures extracted from the literature (171).}
These entries pair molecular drawings with reference coordinates released in the Supporting Information or datasets of 39 publications, including mechanistic studies~\citep{siedschlag2015,wang2025} and structural benchmarks (\Cref{tab:papers}).
They include 
coordination complexes and organometallic catalytic intermediates with diverse ligand arrangements, bonding patterns, and drawing conventions. They extend the benchmark to chemically challenging structures as they appear in the research literature.


\textbf{Validation.}
Each two-dimensional image was manually reviewed alongside its corresponding three-dimensional reference structure.
We checked that the molecular drawing was complete and that atom labels, bonds, charges, and stereochemical annotations were legible.
Reference structures were inspected for agreement with the drawings in atom inventory, connectivity, including metal--ligand bonding, and depicted stereochemical configuration.
This manual curation required interpreting abbreviated groups and implicit hydrogens and tracing stereochemical and coordination cues in each drawing against the reference structure.
This pairwise review checks that the input and the reference describe the same chemical species. It does not require a unique conformation.
Validation criteria are detailed in \Cref{app:validation}.

\subsection{Task formulation}
\label{sec:task}

\textbf{Model input.}
Each system receives an anonymised PNG image and the following user prompt, with the file name changed for each molecule:
\begin{quote}\small\ttfamily\raggedright
A 2D structure of a molecule is provided as a PNG file in your working directory: mol\_0001.png\\[0.5em]
Build the 3D structure it depicts and save it as answer.xyz.
\end{quote}
The accompanying system prompt specifies the assistant's role and output format (\Cref{app:prompt}).
The  images follow standard chemical drawing conventions, including implicit hydrogens, abbreviated groups (e.g. \ce{NMe2}), and solid and hashed wedges reflecting stereochemical information.
Formal charges and, for open-shell species, spin annotations appear on the drawing where applicable.
The input supplies no SMILES string, molecular formula, atom list, reference coordinates, or geometry hints.
Expanding the drawing into a complete atom inventory is part of the task.
The system must save an XYZ file containing Cartesian coordinates for every atom, including hydrogens.
The system prompt used in the experiments said ``Include all explicit hydrogens'', which left the requirement for implicit hydrogens ambiguous (\Cref{app:prompt}).
Image preparation and annotation are described in \Cref{app:depictions}.

\textbf{Agent construction.}
The benchmark accepts any system that reads the input PNG and returns the required XYZ file.
Our implementation uses a lightweight agent: a multimodal large language model (LLM) with four tools.
\texttt{view\_png\_file} supplies the drawing as an image in the model's conversation, so that the model can read it.
\texttt{bash\_kernel} provides a coding workspace in which the model executes commands, constructs coordinates, and writes files.
\texttt{view\_molecule} returns images of the generated structure to the model, with atom labels and viewing angles chosen through tool calls. Rotated views reveal spatial arrangements and overlaps that can be hidden in a single projection.
\texttt{constraint\_xtb} optimises a generated geometry using GFN2-xTB, with optional geometric constraints~\citep{bannwarth2019gfn2}.
The model plans the 2D-to-3D conversion and implements its construction strategy through these tools, principally in the coding workspace. The viewing tools support interpretation and revision.
Tool execution is sandboxed, without access to the internet, reference structures, or answer key (\Cref{sec:agentsetup}).
Other agents, specialised models, and conventional pipelines can follow the same input-PNG/output-XYZ task specification and use the same verifier without adopting this four-tool agent.

\subsubsection*{Verification design}

The verifier first checks that the submitted XYZ file is readable and has the reference element composition.
GFN2-xTB single-point calculations then yield Wiberg bond-order (WBO) matrices for prediction and reference, without relaxing either structure~\citep{bannwarth2019gfn2,bannwarth2021xtb,wiberg1968}.
Both calculations use the charge and spin assigned to the depicted chemical species and recorded with its reference XYZ structure (\Cref{app:charge-spin,app:evaluation}).
Atom-to-atom mapping (AAM) uses MAPPA~\citep{zou2026mappa}, which identifies corresponding atoms in the two weighted graphs, including metal--ligand connections, independently of their numbering.
The molecular-identity check requires a complete correspondence with zero broken or formed bonds under the WBO tolerances.

Using that correspondence, the verifier compares tetrahedral handedness (R/S), supported pyramidal configurations, E/Z relationships, and supported stereogenic axes.
Constraints are defined by distinguishable substituents in the reference, so chemically equivalent hydrogens do not acquire artificial stereochemical identities.
An inversion fails, as does loss of a configuration that the reference defines: a tetrahedral centre that becomes planar in the prediction is rejected rather than skipped.
If this correspondence fails stereochemistry, a symmetry-equivalent correspondence can pass if it preserves the zero bond-event count and satisfies both stereochemistry and ligand geometry.

A separate ligand check verifies reference-defined coordination sites and their angular profiles at metals and supported hypervalent centres.
Reference sites are identified from WBO-supported contacts. At a metal, a site is retained if at least one donor has a WBO decrease of at most 0.50 from reference to prediction. At a supported nonmetal centre, at least one donor must have a predicted WBO of 0.10 or greater. Neither rule uses a bond-length cutoff.
Adjacent donors with $\pi$-bonding support form one $\eta$-bound site, represented by its centroid~\citep{cirera2005shape}. Chemically equivalent site pairs are compared without treating their numbering as significant, and chelate angles are distinguished from angles between separate ligands.
A lost or collapsed site fails. At each checked centre, the RMS difference across corresponding ligand angles must be at most $20^\circ$ and the largest difference at most $60^\circ$.
A pass requires identity, stereochemistry, and ligand geometry to agree. Their numerical thresholds and coverage are specified in \Cref{app:evaluation}.

The task asks for a structure realising the drawing, which usually does not specify a unique conformation.
A valid alternative conformer can have a large whole-molecule RMSD, while averaging deviations across many atoms can obscure a local stereochemical or coordination error.
An RMSD cutoff therefore does not directly establish molecular identity or preserve every defined configuration~\citep{buttenschoen2024posebusters}.
The explicit chemical checks above provide the binary decision without requiring the deposited conformer, a particular energy, or an additional RMSD threshold.

\subsection{Features of \bench{}}
\label{sec:features}

\textbf{Comprehensive chemical coverage.}
The dataset combines 15 local geometry classes, coordination numbers from 2 to 8, and 52 elements, covering organic, hypervalent main-group, and transition-metal chemistry.
VSEPR-selected structures, polycyclic natural products, and literature complexes probe different demands on structure construction, from local geometric interpretation to atom bookkeeping and complex coordination environments.
This breadth exposes deficiencies that a dataset restricted to routine organic molecules may miss.

\textbf{Extendable from the literature.}
New publications can supply both structural drawings and deposited three-dimensional coordinates, allowing additional tasks to be curated by the same pairing and validation procedure.
Such additions can introduce new chemistry, drawing conventions, and more demanding structures as systems improve.
Recent publications can also reduce potential overlap with training data.
Retaining source identifiers and documenting the benchmark release used for each score keeps evaluations traceable as the set grows.

\textbf{A general prompt and freedom of construction.}
The prompt specifies the input and required output without prescribing a construction method or supplying a scoring strategy.
Within the declared environment and budget, a system may use SMILES, connection tables, direct interpretation of the image, coordinate generation, or any combination of these approaches.
Intermediate representations do not affect the score. Only the submitted structure is evaluated.

\textbf{Consistent verification across systems.}
Every prediction is processed by the same WBO, AAM, stereochemistry, and ligand-geometry checks, independently of the system or construction route that produced it.
Atom mapping accommodates arbitrary atom order, symmetry handling avoids penalising equivalent relabellings, and the bond-order graphs include coordination bonding.
This supplies a common automated measure of molecular identity, the checked stereochemical configurations, and coarse ligand geometry across chemically diverse inputs.

\textbf{Conformer freedom.}
Molecular identity is determined from WBO-based bonding after MAPPA atom mapping, including metal--ligand connections.
RMSD instead measures geometric similarity to a particular reference conformation: a correct alternative conformer can have a large RMSD, while averaging over all atoms can dilute a local structural error.
Low RMSD alone therefore does not establish chemical correctness~\citep{buttenschoen2024posebusters}.
Although RMSD can be thresholded to give a binary verdict, a geometric cutoff does not directly test molecular identity.
\bench{} makes its binary decision using composition, WBO-based connectivity, stereochemistry, and coarse ligand geometry, with explicit numerical tolerances (\Cref{sec:task}).
Any conformer passing these checks is accepted irrespective of its relative energy. Reproducing the reference conformer or finding a lowest-energy geometry is not required.
By accepting alternative conformations, \bench{} isolates the ability to interpret a chemical drawing and construct the corresponding molecule in three dimensions.

\textbf{Diagnosable failures.}
Geometry, size, and organic structural complexity annotations allow resolve rates to be analysed over chemically meaningful subsets.
Separate failure categories for missing output, parsing, composition, connectivity, stereochemistry, and ligand geometry distinguish an incorrect atom inventory from errors in bonding or stereocentre configuration.

\section{Evaluation setup}
\label{sec:setup}
\label{sec:systems}

\textbf{Model selection.}
\label{sec:models}
\label{sec:agents}
We test nine multimodal LLMs from OpenAI, Anthropic, Alibaba, and Moonshot AI (\Cref{tab:models}).
Reasoning effort is high for all models except qwen3.8-max, which uses medium.
The Qwen interface offers low, medium, and xhigh but no high setting, so we select medium as the most suitable option for this comparison.

\begin{table}[H]
  \centering
  \captionsetup{skip=4pt}
  \caption{\textbf{LLMs used in the agent.} Exact API model identifiers, developers, and reasoning effort for the nine tested models.}
  \label{tab:models}
  \small
  \begin{tabular}{@{}llc@{}}
  \toprule
  Model identifier & Developer & Reasoning effort \\
  \midrule
  \texttt{gpt-6.1-sol}~\citep{openai2026sol61}      & OpenAI      & high \\
  \texttt{gpt-6-astra}~\citep{openai2026astra}      & OpenAI      & high \\
  \texttt{gpt-6-sol}~\citep{openai2026sol}          & OpenAI      & high \\
  \texttt{gpt-6-luna}~\citep{openai2026luna}       & OpenAI      & high \\
  \texttt{claude-fable-5-1}~\citep{anthropic2026fable51} & Anthropic & high \\
  \texttt{claude-opus-5-5}~\citep{anthropic2026opus55} & Anthropic & high \\
  \texttt{claude-opus-5}~\citep{anthropic2026opus5}  & Anthropic   & high \\
  \texttt{qwen3.8-max}~\citep{qwen2026max38}       & Alibaba     & medium \\
  \texttt{kimi-k3}~\citep{moonshot2026kimi3}       & Moonshot AI & high \\
  \bottomrule
  \end{tabular}
\end{table}

\textbf{Environment and input.}
\label{sec:agentsetup}
\label{sec:input}
\label{sec:prompt}
We use a reduced four-tool configuration of {\cinzel El Agente Estructural}~\citep{choi2026estructural}.
The four tools, system prompt, execution environment, and budget of 100 model calls per molecule are shared across runs.
Tool calls are issued one at a time, and each molecule contributes one scored session per model to the main comparison (\Cref{app:execution-budget}).
Every shell command runs in its own user, mount, and network namespace, with access to the anonymised PNG, the working directory, and installed software. The internet, the reference geometries, and the answer key are out of reach.
The input condition and user prompt are defined in \Cref{sec:task}, with the system prompt in \Cref{app:prompt}.
A session that reaches the 100-model-call limit without returning a structure is scored as no answer.
For gpt-6.1-sol, we additionally compare the four-tool configuration with only \texttt{view\_png\_file} and \texttt{bash\_kernel}, preserving access to the drawing and a flexible coding workspace.
The model, reasoning effort, prompts, 300 molecules, budget, sandbox, and verifier are held fixed to assess the contribution of the dedicated structure-viewing and geometry-optimisation tools.
Ablation results and resource use are reported in \Cref{app:ablation} and \Cref{fig:tool-ablation}. The full evaluation setup is given in \Cref{app:setup}.

\section{Results}
\label{sec:results}

We compare accuracy and resource use for nine models over 300 molecules, then examine how performance varies with molecular atom count and chemical category.

\subsection{Accuracy and resource use}
\label{sec:res-systems}

\begin{figure}[H]
    \centering
    \captionsetup{skip=4pt}
    \includegraphics[width=0.94\textwidth]{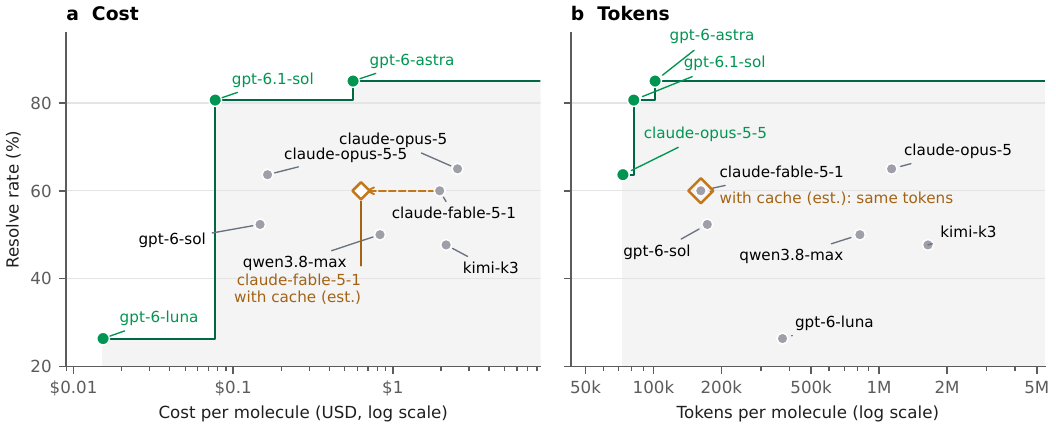}
    \caption{\textbf{Accuracy against resources for nine models.} \emph{Left:} cost per molecule. \emph{Right:} total input plus output tokens per molecule. Passes, total cost, and tokens are each divided by all 300 attempts, including verifier errors and missing answers. Green points, model labels, and lines identify the Pareto frontiers; grey points are dominated. Thin connectors in both panels link labels to their corresponding points. Orange diamonds mark the Fable 5.1 caching estimate using Opus 5.5's cache-read and cache-write proportions; token counts and accuracy are unchanged, so the token marker overlaps the observed point.}
    \label{fig:cost-tokens}
\end{figure}

\textbf{Pareto frontiers.}
We compare reconstruction success against monetary cost and, separately, token use, through Pareto frontiers (\Cref{fig:cost-tokens}).
A model lies on a frontier if no other tested model is at least as accurate and uses no more resources, with a strict improvement in at least one of these measures.
For \bench{}, these frontiers identify efficient choices under different resource budgets and show the additional cost or token use needed to achieve higher success rates.

\textbf{Cost per molecule.}
Overall resolve rates range from 26.3\% to 85.0\%, with every model evaluated on all 300 molecules (\Cref{fig:cost-tokens} and \Cref{tab:efficiency}).
The cost--accuracy Pareto frontier comprises gpt-6-luna (26.3\%, \$0.0153 per molecule), gpt-6.1-sol (80.7\%, \$0.0771), and gpt-6-astra (85.0\%, \$0.5634).
gpt-6.1-sol therefore achieves a rate 4.3 percentage points below gpt-6-astra at about one seventh of its cost.
The more expensive claude-opus-5, kimi-k3, and claude-fable-5-1 runs cost \$2.5431, \$2.1568, and \$1.9657 per molecule while resolving 65.0\%, 47.7\%, and 60.0\%, respectively. Each is dominated by a cheaper, more accurate configuration.
claude-fable-5-1 recorded no cache hits. Applying claude-opus-5-5's cache-read and cache-write fractions of input tokens (91.27\% and 8.70\%, respectively) gives an estimated cost of \$0.6325 per molecule, with its observed token counts and 60.0\% resolve rate unchanged.

\textbf{Tokens per molecule.}
The token--accuracy Pareto frontier comprises claude-opus-5-5 (63.7\%, 73.2k tokens per molecule), gpt-6.1-sol (80.7\%, 81.8k), and gpt-6-astra (85.0\%, 101.7k).
By comparison, claude-opus-5 and kimi-k3 use 1.14M and 1.65M tokens per molecule.
The Fable 5.1 caching scenario retains 162.2k total tokens per molecule, so its marker coincides with the observed result in the token panel.
Higher token consumption therefore does not consistently yield higher accuracy in these runs.
The two frontiers identify different trade-offs: gpt-6-luna minimizes monetary cost, claude-opus-5-5 minimizes token use, and gpt-6.1-sol and gpt-6-astra trade additional resources for higher success rates.
All nine models' values are reported in \Cref{tab:efficiency}.

\subsection{Difficulty correlates with molecular atom count}
\label{sec:res-size}

\begin{figure}[H]
    \centering
    \captionsetup{skip=4pt}
    \includegraphics[width=\textwidth]{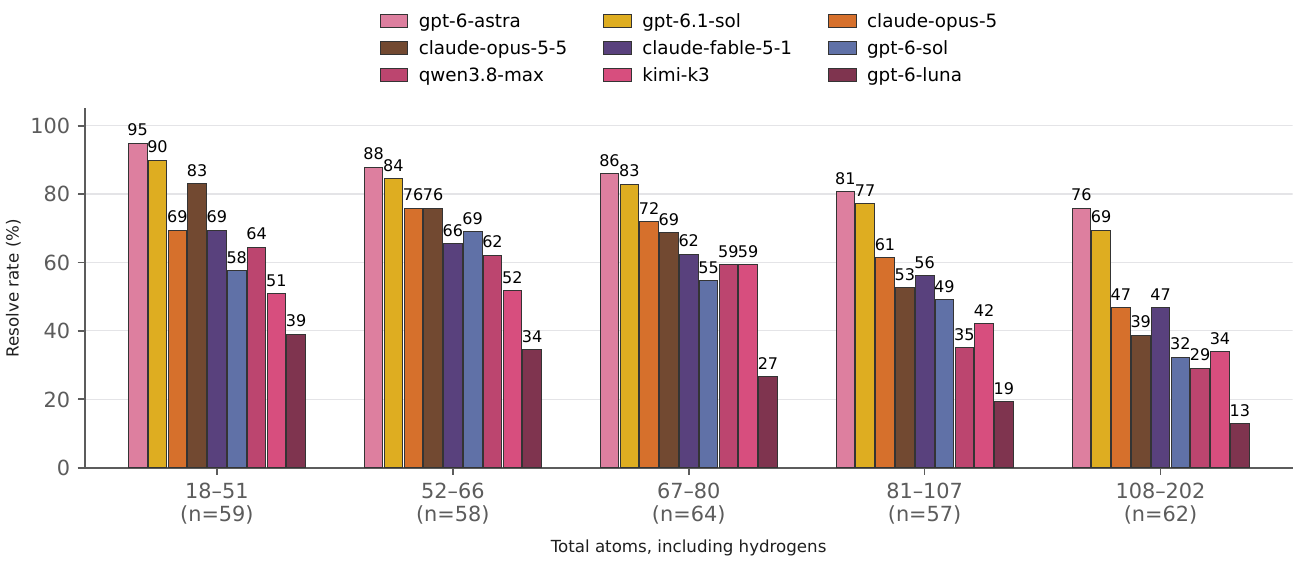}
    \caption{\textbf{Success against total atom count, including hydrogens.} Each bin shows the resolve rate of all nine models in legend order, with its molecule count below the axis. Values above bars are percentages rounded to the nearest integer. Each rate uses every molecule in its bin, including verifier errors and missing answers.}
    \label{fig:atom-count}
\end{figure}

\textbf{Success decreases with total atom count.}
This trend is measured on a deliberately demanding molecular set: the natural products were selected for high structural complexity, and the literature entries include large, diverse coordination compounds rather than a random sample of routine organics.
Reference structures contain 18--202 atoms, including hydrogens.
Across atom-count quintiles of 18--51, 52--66, 67--80, 81--107, and 108--202 atoms, mean resolve rates decrease from 68.7\% through 67.4\%, 63.5\%, and 52.6\% to 42.8\% (\Cref{fig:atom-count} and \Cref{tab:atom-difficulty}).
At the molecule level, total atom count has a negative Spearman correlation with the fraction of models resolving that molecule ($\rho=-0.332$, $n=300$ molecules).
Larger molecules therefore tend to be harder in this benchmark, although the trend does not imply that every larger molecule is harder than every smaller one.
Chemical category and structural complexity also vary with size, so the association does not isolate an effect of atom count alone.

\subsection{Difficulty varies by chemical category}
\label{sec:res-source}

\textbf{VSEPR-curated, natural-product, and literature structures.}
Mean resolve rates across the nine models are 69.9\% for VSEPR-curated structures, 69.2\% for natural products, and 51.0\% for literature structures (\Cref{fig:by-source} and \Cref{tab:results}).
The literature group is the lowest-scoring source for all nine models. It contains 156 metal-bearing structures out of 171 and has a median of 46 heavy atoms, compared with 35 for VSEPR-curated structures and 28 for natural products (\Cref{tab:stats}).
Its lower success is consistent with the demands of larger coordination complexes, varied metal--ligand arrangements, and diverse drawing conventions.
VSEPR-curated entries provide focused tests of local geometry with uniform depictions, while natural products mainly test organic connectivity and multiple stereocentres.
These design differences help interpret the source-level pattern, but their contributions cannot be separated by this comparison alone.
Within-category analyses are reported in \Cref{app:category-difficulty}.

\begingroup
\setlength{\intextsep}{4pt}
\begin{figure}[H]
    \centering
    \captionsetup{skip=4pt}
    \includegraphics[width=0.78\textwidth]{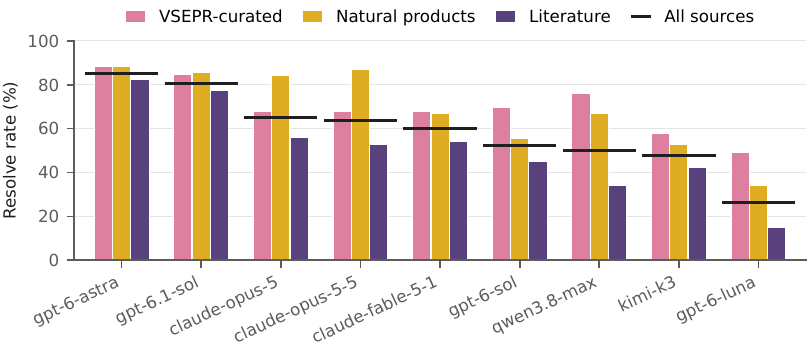}
    \caption{\textbf{Resolve rates across the three chemical categories.} Bars show each model's rates on VSEPR-curated structures ($n=59$), natural products ($n=70$), and literature structures ($n=171$); black lines show overall rates. Each category rate uses its full subset size, with verifier errors retained. Overall rates use 300 attempts.}
    \label{fig:by-source}
\end{figure}
\endgroup

\subsection{Qualitative analysis}
\label{sec:qualitative}

We examine a failed reconstruction by gpt-6.1-sol on a molecule that another model also fails to reconstruct.
Failure mechanisms can differ across models even for the same molecule.
This case provides insight into a specific reconstruction error, without implying that it represents the failure modes of all models.

\begin{figure}[H]
    \centering
    \captionsetup{skip=4pt}
    \includegraphics[width=0.80\textwidth]{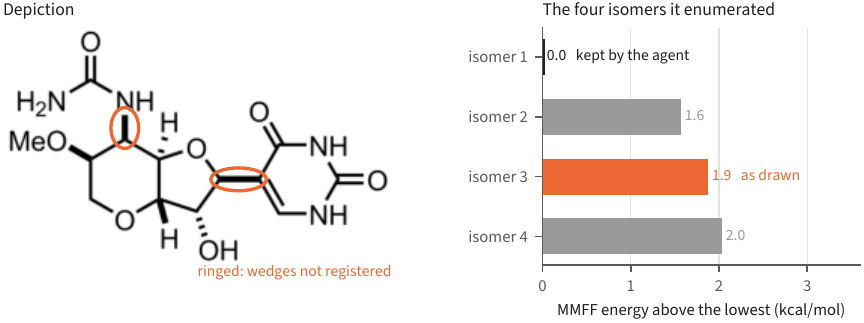}
    \caption{\textbf{gpt-6.1-sol on malayamycin~A.} \emph{Left:} the drawing. Orange ovals highlight the two bold bonds to the urea nitrogen and uracil ring whose stereochemical cues the agent's build script failed to encode. \emph{Right:} the four stereoisomers it enumerated for the two centres it left open, at the lowest MMFF energy of each, relative to the lowest, as printed in the session. It kept the lowest (black), which differs from the isomer the drawing shows (orange) at the C-glycosidic carbon.}
    \label{fig:qualitative}
\end{figure}

\textbf{An inverted centre can come from a drawing read almost right, and an energy used to decide what the drawing had fixed.}
\Cref{fig:qualitative} follows gpt-6.1-sol on malayamycin~A, a natural product with six stereocentres that seven of the nine LLMs resolve.
The agent viewed the drawing and one enlarged crop of it, then wrote a single script that builds the molecule atom by atom in RDKit from the pixel coordinates of the drawing, with each wedge and hash carried as a bond direction; the session took seven tool calls.
The script registered the wedges at four of the six centres but not the two bold bonds to the urea nitrogen and to the uracil ring, and its own comment shows how it closed the gap: ``\textit{Sample either configuration there, and retain the lowest-energy conformer}.''
It enumerated the four stereoisomers of those two centres, embedded up to fifty conformers of each, and kept the lowest by MMFF. The four lie within 2.0~kcal/mol, and the one it kept lies 1.9~kcal/mol below the isomer the drawing shows and differs from it at the C-glycosidic carbon.
After a GFN2-xTB optimisation the structure passed the identity gate and failed the stereochemistry gate on that one centre of six, while gpt-6-astra, reading the same drawing, set all six (``solid wedges/bold exocyclic bonds are forward'').
The error is not a misunderstanding of what a wedge means but the use of an energy, which cannot decide a configuration, in place of a cue the drawing gives. \Cref{app:qualitative} walks through ten further sessions, covering the other ways the systems fail and recover.

\section{Related work}
\label{sec:related}

\textbf{Benchmarks for language models in chemistry.}
Chemical competence has been assessed through exam-style knowledge and reasoning questions~\citep{mirza2025chembench}, short-answer molecular comprehension~\citep{runcie2025chemiq}, and adaptations of classical property-, reaction-, and retrosynthesis-prediction tasks to a prompting setting~\citep{guo2023chemllmbench}.
Multimodal benchmarks extend the input to figures, spectra, and laboratory images: MaCBench evaluates vision-language models across data extraction, experimental understanding, and results interpretation~\citep{alampara2025macbench}, MolPuzzle asks models to elucidate structures from spectra~\citep{guo2024molpuzzle}, and ChemVLM is trained and evaluated on chemical images including structural diagrams~\citep{li2024chemvlm}.
ScienceAgentBench evaluates generated programs and their execution on scientific data-analysis tasks~\citep{chen2024scienceagentbench}, while LAB-Bench uses multiple-choice questions to assess practical biology-research capabilities~\citep{laurent2024labbench}.
\bench{} focuses on reconstructing a complete molecular structure from a chemical drawing.

\textbf{Three-dimensional structure generation with language models.}
Optical chemical structure recognition recovers molecular graphs from drawings~\citep{qian2023molscribe,krasnov2024comparing}, while conformer-generation workflows construct three-dimensional coordinates from a specified molecular graph~\citep{riniker2015etkdg,demuth2026strained}.
Language models can emit molecular geometries, crystal structures, and protein binding sites directly as coordinate files~\citep{flamshepherd2023xyz}, can be coupled to three-dimensional encoders for molecule--text tasks~\citep{li2024molm3d}, and, for transition-metal complexes specifically, can be given a geometry-aware encoder trained on large complex libraries~\citep{zhu2026tmcllm}.
Fine-tuning a language model on Cartesian coordinates or Z-matrices yields equilibrium geometries and conformer ensembles competitive with specialised models, while frontier models without fine-tuning perform poorly on the same task~\citep{cavanagh2026geometry}.
Related benchmarks evaluate spatially constrained ligand generation for structure-based drug design~\citep{macdougall2026spatial}, and PoseBusters establishes that generated three-dimensional structures must be checked for physical validity rather than scored on geometric deviation alone~\citep{buttenschoen2024posebusters}.
These methods differ in their input representations, chemical coverage, and training objectives.
\bench{} evaluates drawing-to-coordinate conversion across organic, main-group, and coordination chemistry under a common input and output specification.

\textbf{Spatial reasoning and structure manipulation by agents.}
{\cinzel El Agente Estructural} supports natural-language-guided generation and editing of molecular geometries~\citep{choi2026estructural}, while {\cinzel El Agente Q} plans, executes, and analyses quantum-chemical workflows~\citep{zou2025elagente}.
AtomWorld benchmarks spatial reasoning in language models through structure-modification tasks on crystallographic files, finding that success falls sharply as the spatial complexity of the operation increases~\citep{lv2025atomworld}.
\bench{} instead assesses construction from two-dimensional molecular drawings, including the recovery of atom connectivity and stereochemical constraints before a complete coordinate file is produced.

\section{Outlook}
\label{sec:discussion}

Expanding \bench{} beyond the current 300 molecules would allow more detailed analyses within each molecular category.
Larger samples spanning a wider range of sizes, organic frameworks, and coordination environments would help establish whether the observed decline in success with atom count also holds within individual categories, and how difficulty trends differ between them.

Controlling input-image resolution, dimensions, and legibility would make model comparisons more interpretable.
Evaluating matched depictions at standardised quality, followed by controlled changes in resolution, would help separate sensitivity to image presentation from molecular reconstruction difficulty.

\bench{} can also support comparisons among a wider range of 2D-to-3D systems.
Conventional pipelines combining optical chemical structure recognition with three-dimensional structure generation, specialised conversion models, and hybrid approaches could be evaluated alongside LLM agents.
Using the same input drawings, output requirements, and verification criteria would provide a common basis for comparing these approaches and determining which methods handle different chemical classes most reliably.

The paired data could also support a reverse task in which a system generates a two-dimensional chemical drawing from three-dimensional coordinates.
Evaluation would assess preservation of connectivity and stereochemical information while allowing alternative drawing layouts.

A further direction is to assess whether generated structures provide usable starting geometries for downstream calculations, including geometry optimisation, molecular docking, and quantum-chemical workflows.
Application-specific measures, such as optimisation convergence and the structural changes required during relaxation, would complement the existing checks of molecular identity, stereochemistry, and coarse ligand geometry.
This additional assessment would connect the successful reconstruction of the depicted molecule with its practical use in computational chemistry.

\section{Conclusions}
A chemical drawing fixes which atoms are bonded and how each stereocentre is configured, and leaves the shape to the reader. Language models can now recover many three-dimensional structures from such drawings, but not yet reliably (\Cref{sec:res-systems}). Their failures followed the chemistry. Success fell for the largest molecules (\Cref{sec:res-size}) and was lowest for structures from the literature, most of them coordination complexes with varied metal--ligand arrangements and drawing conventions (\Cref{sec:res-source}). Higher cost and token use did not consistently raise accuracy. The present set is too small to separate the effect of size from that of chemistry.

For organic molecules of moderate size, an agent-built structure is a reasonable starting point. For metal complexes, each coordination site and its geometry should be checked against the drawing (\Cref{sec:res-source}). For any molecule with defined stereochemistry, each stereocentre should be checked as well (\Cref{sec:qualitative}). For example, the agent missed two bold bonds of malayamycin~A, assigned their configurations by conformer energy, and inverted the C-glycosidic carbon.

\bench{} scores bonding, configuration, and coordination, and accepts any conformer that preserves them. The same verifier can therefore compare conventional pipelines, specialised models, and agents on one set, and new structures from the literature can extend it. We expect coordination complexes to separate these systems most clearly, because they were the hardest structures for every model tested here.


\clearpage
\section*{Conflicts of interest}
There are no conflicts of interest to declare.

\section*{Acknowledgments}
We gratefully acknowledge the longstanding contributions of the Matter Lab's current and past group members (\url{https://matter.toronto.edu}), in particular the El Agente subgroup.

O.N.F. acknowledges support from the Spanish Ministry of Science, Innovation and Universities (MICIU) through a Salvador de Madariaga research mobility fellowship (grant PRX24/00448).

A.A.-G. thanks Anders~G.~Fr{\o}seth for his generous support. A.A.-G. also acknowledges the generous support of Natural Resources Canada and the Canada 150 Research Chairs program.
\acknowAC

\acknowDARPA

This work was supported by the AI2050 program of Schmidt Sciences.

This research was enabled in part by support provided by the SciNet HPC Consortium (\href{https://scinethpc.ca}{scinethpc.ca}), Calcul Qu\'ebec (\href{https://www.calculquebec.ca}{www.calculquebec.ca}), and the Digital Research Alliance of Canada (\href{https://alliancecan.ca}{alliancecan.ca}).

\clearpage


{
\small
\bibliography{references,references-data,references-methods}
\bibliographystyle{unsrtnat}   
}


\clearpage

\appendix


\section{Benchmark details}
\label{app:benchmark}
\setlength{\intextsep}{6pt plus 2pt minus 2pt}
\setlength{\textfloatsep}{10pt plus 2pt minus 2pt}

\subsection{Sources, identifiers, and licences}
\label{app:sources}
Each entry carries the identifier of its source record---a PubChem compound identifier, a COD entry number, or a CSD refcode---or, for entries taken from the literature, the DOI of the paper and the label the paper gives the species.
\Cref{tab:sources} lists, for each part of the benchmark, where the reference coordinates and the depiction come from and the references that document them; \Cref{tab:papers} breaks the literature source down by publication.
The primary publications of the crystal structures behind the VSEPR-curated entries are cited as~\citeDataCODPrimary{} for the 17 entries cut from COD structures and as~\citeDataCSDPrimary{} for the 41 CSD structures curated in tmQM.
The reference coordinates come from several sources: PubChem, on whose data NCBI places no restriction of its own; PubChemQC, released under CC BY 4.0; the COD, dedicated to the public domain under CC0; and the Supporting Information and data sets of the cited papers, of which the MOR41 repository declares no licence.
The 41 transition-metal geometries are tmQM's, whose repository carries the MIT licence; they derive from CSD structures.
Depictions redrawn from reference coordinates are our own.
All third-party material reproduced in this article is covered by its source licence or by permission obtained for publication in the article.
Among the 159 literature depictions cropped from published figures, 35 have confirmed CC BY sources (permitting reuse with attribution, including commercial use), including Paterson et al.~\citep{paterson2020}, and 45 come from Supporting Information deposited under CC BY-NC 4.0 (permitting reuse with attribution for noncommercial purposes only), including five DU8+ drawings~\citep{novitskiy2021} and 22 ROST61 drawings~\citep{maurer2021}.
Electronic republication permissions for 19 ACS source drawings have been obtained through Copyright Clearance Center; their scope and credit notices are detailed below.
The ten edited drawings from Anderson et al.~\citep{anderson2024} are reproduced here with permission.
The 70 natural-product depictions are SynArchive's target drawings~\citep{synarchive}.

\begin{table}[!htbp]
  \centering
  \caption{\textbf{Provenance of \bench{}.} Where the reference coordinates and the depiction of each part of the benchmark come from. ``Drawn'' depictions were produced by us from the reference coordinates (\Cref{app:depictions}).}
  \label{tab:sources}
  \small
  \renewcommand{\arraystretch}{1.3}
  \begin{tabular}{@{}lr>{\raggedright\arraybackslash}p{5.2cm}>{\raggedright\arraybackslash}p{3.6cm}@{}}
  \toprule
  Source & $n$ & Reference coordinates & Depiction \\
  \midrule
  VSEPR-curated, organic & 1 & PubChem 3D conformer~\citep{kim2025pubchem} & drawn \\
  VSEPR-curated, from COD & 17 & molecule or fragment of a COD crystal structure~\citep{grazulis2009,grazulis2012cod} & drawn \\
  VSEPR-curated, transition metal & 41 & tmQM GFN2-xTB geometry of a CSD structure~\citep{balcells2020tmqm,groom2016csd} & drawn \\
  Natural products & 70 & PubChem (61), PubChemQC-PM6 (8)~\citep{kim2025pubchem,nakata2020}, or wwPDB (1) & SynArchive target drawing~\citep{synarchive} \\
  Literature & 171 & Supporting Information or dataset of the paper (\Cref{tab:papers}) & figure from the paper or drawing from its coordinates \\
  \midrule
  Total & 300 & & \\
  \bottomrule
  \end{tabular}
\end{table}

\FloatBarrier
\begingroup
\small
\renewcommand{\arraystretch}{1.15}
\begin{longtable}{@{}llr@{}}
  \caption{\textbf{Literature sources of \bench{}.} Number of molecules supplied by each publication to the combined literature source.}\label{tab:papers}\\
  \toprule
  Source & Journal & Molecules \\
  \midrule
  \endfirsthead
  \multicolumn{3}{l}{\small\textbf{Table \thetable{} (continued)}}\\
  \toprule
  Source & Journal & Molecules \\
  \midrule
  \endhead
  \midrule
  \multicolumn{3}{r}{\small Continued on next page}\\
  \endfoot
  \bottomrule
  \endlastfoot
  Zhang et al.\ (2017)~\citep{zhang2017} & J. Am. Chem. Soc. & 10 \\
  Anderson et al.\ (2024)~\citep{anderson2024} & Nat. Commun. & 10 \\
  Siedschlag et al.\ (2015)~\citep{siedschlag2015} & J. Am. Chem. Soc. & 5 \\
  Wang et al.\ (2025)~\citep{wang2025} & J. Org. Chem. & 7 \\
  Paterson et al.\ (2020)~\citep{paterson2020} & Org. Lett. & 2 \\
  DU8, Novitskiy et al.\ (2021)~\citep{novitskiy2021} & J. Org. Chem. & 5 \\
  MOR41, Dohm et al.\ (2018)~\citep{dohm2018} & J. Chem. Theory Comput. & 9 \\
  ROST61, Maurer et al.\ (2021)~\citep{maurer2021} & J. Chem. Theory Comput. & 22 \\
  PMC12458972, Zhou et al.\ (2025)~\citep{zhou2025} & Organometallics & 3 \\
  PMC12892316, Feresin et al.\ (2026)~\citep{feresin2026} & Inorg. Chem. & 3 \\
  PMC12869483, Jayasundara et al.\ (2026)~\citep{jayasundara2026} & Inorg. Chem. & 2 \\
  PMC12977057, Button-Jennings et al.\ (2026)~\citep{buttonjennings2026} & Organometallics & 1 \\
  PMC12326349, Stephan et al.\ (2025)~\citep{stephan2025} & Inorg. Chem. & 6 \\
  PMC12308804, Sledesky et al.\ (2025)~\citep{sledesky2025} & Inorg. Chem. & 2 \\
  PMC12522133, Sittel et al.\ (2025)~\citep{sittel2025} & Inorg. Chem. & 1 \\
  PMC12690572, Bajo et al.\ (2025)~\citep{bajo2025} & Organometallics & 4 \\
  PMC12334357, Yang et al.\ (2025)~\citep{yang2025} & Nat. Synth. & 2 \\
  PMC13058889, Nieto-Vargas et al.\ (2026)~\citep{nietovargas2026} & Inorg. Chem. & 4 \\
  PMC12869501, Morganti et al.\ (2026)~\citep{morganti2026} & Inorg. Chem. & 2 \\
  PMC13580183, Melnikov et al.\ (2026)~\citep{melnikov2026} & Organometallics & 7 \\
  PMC12076555, Shiels et al.\ (2025)~\citep{shiels2025} & Inorg. Chem. & 1 \\
  PMC13390040, Padilla et al.\ (2026)~\citep{padilla2026} & Inorg. Chem. & 8 \\
  PMC12930352, Gross et al.\ (2026)~\citep{gross2026} & ACS Catal. & 7 \\
  PMC13418093, Pita-Milleiro et al.\ (2026)~\citep{pitamilleiro2026} & Inorg. Chem. & 8 \\
  PMC11894599, Ferrer-Bru et al.\ (2025)~\citep{ferrerbru2025} & Inorg. Chem. & 7 \\
  PMC13231423, García-Vega et al.\ (2026)~\citep{garciavega2026} & Inorg. Chem. & 3 \\
  PMC12933881, Webber et al.\ (2026)~\citep{webber2026} & Organometallics & 1 \\
  PMC12820928, Walsh et al.\ (2026)~\citep{walsh2026} & Inorg. Chem. & 8 \\
  PMC13081111, Murray et al.\ (2026)~\citep{murray2026} & Organometallics & 1 \\
  BJOC\_12\_13, Poater et al.\ (2016)~\citep{poater2016} & Beilstein J. Org. Chem. & 3 \\
  PMC12856905, Díaz-Jiménez et al.\ (2026)~\citep{diazjimenez2026} & J. Am. Chem. Soc. & 2 \\
  PMC12522683, Benzan Lantigua et al.\ (2025)~\citep{benzanlantigua2025} & Organometallics & 2 \\
  PMC12818750, Figueira et al.\ (2025)~\citep{figueira2025} & Inorg. Chem. & 1 \\
  PMC12153055, Pagliaricci et al.\ (2025)~\citep{pagliaricci2025} & Organometallics & 4 \\
  PMC12606712, Siano et al.\ (2025)~\citep{siano2025} & Organometallics & 3 \\
  PMC12709582, Morello et al.\ (2025)~\citep{morello2025} & Inorg. Chem. & 2 \\
  PMC13371084, Seith et al.\ (2026)~\citep{seith2026} & Organometallics & 1 \\
  PMC12235671, Bermejo et al.\ (2025)~\citep{bermejo2025} & ACS Catal. & 1 \\
  PMC9554919, Isaac et al.\ (2022)~\citep{isaac2022} & Organometallics & 1 \\
  \midrule
  Total & & 171 \\
\end{longtable}
\endgroup

\textbf{ACS image sources and copyright notices.}
The nine files with prefix \texttt{shot\_mor41\_} are individual crops from Figure~1 of Dohm et al.~\citep{dohm2018}.
Their original species labels are PR25, PR16, PR41, ED26, PR36, PR22, PR04, PR10, and PR24.
The structures were re-cropped at 600 dpi.
Used with permission of the American Chemical Society, from \emph{Comprehensive Thermochemical Benchmark Set of Realistic Closed-Shell Metal Organic Reactions}, Dohm et al., \emph{J. Chem. Theory Comput.} 2018, 14, 2596--2608, \url{https://doi.org/10.1021/acs.jctc.7b01183}; permission conveyed through Copyright Clearance Center, Inc.
Copyright 2018 American Chemical Society. Licence 1815304-1.

The ten files with prefix \texttt{shot\_paper3\_} contain compounds 9, 16, 17, 20, 21, 23, 24, 26, 27, and 31 from Zhang et al.~\citep{zhang2017}.
In addition to cropping, the recorded edits remove selected compound numbers, convert coloured bonds to black, and move the overlapping 2-Nap label in compound~9 while preserving the depicted configuration.
Used with permission of the American Chemical Society, from \emph{Mechanism and Origins of Ligand-Controlled Stereoselectivity of Ni-Catalyzed Suzuki--Miyaura Coupling with Benzylic Esters: A Computational Study}, Zhang et al., \emph{J. Am. Chem. Soc.} 2017, 139, 12994--13005, \url{https://doi.org/10.1021/jacs.7b04973}; permission conveyed through Copyright Clearance Center, Inc.
Copyright 2017 American Chemical Society. Licence 1815304-2.

\FloatBarrier
\subsection{Construction process}
\label{app:construction}
The released manifest carries, for each entry, its name, source, group, original identifier, molecular formula, and atom count; a reference table carries the origin of its coordinates, its database identifier, and the origin of its depiction, including the SynArchive page of each natural product and every edit made to a picture; the accompanying geometry and descriptor tables carry the defining centre, its coordination number and geometry class, the full distribution of classes over every centre in the molecule, the net charge and spin multiplicity, and the size and complexity descriptors of \Cref{app:classification}.
The entries of each source were chosen as follows.

\textbf{VSEPR-curated.}
Candidates were assembled separately for each local-geometry class, defined by the number and spatial arrangement of coordination sites: organic centres from PubChem, hypervalent main-group centres from COD, and transition-metal centres from tmQM.
Candidates with defective coordinates were excluded---one lacking its hydrogens, one carrying the hydrogens of a lattice water on a carbonyl oxygen---as was one whose picture and coordinates turned out to be of different species.
The eight simplest of the rest were then dropped, ranked first by a geometry level (0 for tetrahedral and bent; 1 for linear, trigonal planar, trigonal pyramidal, and octahedral; 2 for square planar, square pyramidal, and trigonal bipyramidal; 3 for seesaw and T-shaped; 4 for pentagonal bipyramidal; one level more at a hypervalent main-group centre) and then by spacial score, except where dropping an entry would have left its class empty, so that all twelve classes the candidates realise are kept.

\textbf{Natural products.}
We selected the 70 SynArchive targets with the highest spacial score (SPS), using SPS as an index of molecular structural complexity~\citep{krzyzanowski2023sps}.
Compound identifiers are taken from the header of the reference file where present and otherwise resolved from the compound name through PubChem.

\textbf{Literature structures.}
The 171 molecules were selected from papers dealing with mechanistic studies, natural-product structure reassignment, transition-metal reaction-energy benchmarks, and papers reporting organometallic catalysts and coordination complexes (\Cref{tab:papers}).
Each entry pairs a 2D molecular depiction with coordinates released by the same study, matched using the species identifier supplied by its authors.
For the collection from 31 open-access organometallic papers, the 101 candidates with the highest raw SPS among the eligible candidates were retained.
This was a selection heuristic for that collection, not an $n$SPS-based selection of all 171 literature structures or a validated complexity measure for metal coordination.
The subsequent $n$SPS analysis is restricted to metal-free organic molecules, including 15 literature entries, and is separate from this selection step (\Cref{app:classification}).
Candidates were excluded where the coordinates and drawing represent different species or configurations, or where a required figure could not be redistributed under the selection criteria.
The literature source is analysed as one group irrespective of how its depiction was obtained.

These subsets were chosen to test complementary challenges: diverse local geometries in organic, hypervalent main-group, and transition-metal species; complex ring systems and stereochemistry in natural products; and the interpretation of published depictions of catalysts, intermediates, and coordination complexes.
We deliberately favoured structurally complex molecules to provide a demanding test of agents' ability to construct chemically faithful three-dimensional structures from two-dimensional drawings.

\subsection{Entries taken from publications}
\label{app:literature}
The 171 literature entries come from 39 publications (\Cref{tab:papers}), two of which define benchmark sets of transition-metal reaction energies.
MOR41 coordinates were taken from the data repository of that benchmark set~\citep{grimmelab}, and its depictions are crops of Figure~1 of the MOR41 paper~\citep{dohm2018}; ROST61 coordinates come from the structure archive released with that paper~\citep{maurer2021}, and its depictions are crops of the reaction schemes in its Supporting Information; both were re-cropped at 600 dpi.
Every other literature entry takes the coordinates the paper's Supporting Information deposits under the label the paper gives the species.
The spin of the 22 open-shell ROST61 molecules is read from the unrestricted calculations of the ROST61 archive, that of the five open-shell dicobalt species from the structure files their coordinates came from, and that of the seven open-shell organometallic complexes from their papers.
No literature entry shares its molecular formula with a VSEPR-curated entry.

\subsection{Depiction generation}
\label{app:depictions}
For VSEPR-curated depictions and literature depictions redrawn from coordinates, an automated script converted reference XYZ coordinates into CDXML drawings, which were rendered in ChemDraw~\citep{evans2014chemdraw}.
Metal--ligand bonds are drawn as plain lines, without dative arrows; substituents are condensed as a chemist would write them, such as \ce{(tBu)2P}; and wedge and hash bonds are drawn where the arrangement at a centre is three-dimensional and omitted where a flat drawing is unambiguous.
The 70 natural-product depictions are SynArchive's drawings of the synthetic targets~\citep{synarchive}; every one of the 70 names matches a SynArchive entry.
One of them shows a chloride counter-ion that the reference does not contain; it was erased, and the ammonium charge it balanced stays drawn.
For literature depictions reproduced from published figures, the figure was cropped to the species and edited only where the crop would otherwise mislead: compound numbers, NMR locants, and colour highlights were removed; counter-ions absent from the reference coordinates were erased; an icon defined only in a caption was replaced by the group it stands for or a standard abbreviation; a generic metal ``M'' in one ROST61 drawing was relabelled Ru, as the caption in the Supporting Information specifies; and in one figure a label overlapping a carbonyl oxygen was moved, its hashed bond redrawn in the same configuration.

\textbf{Charge and spin.}
\label{app:charge-spin}
Here the reference charge and spin mean the net charge and spin assignment of the depicted species, stored with its deposited reference geometry; they are not inferred from the model's submitted coordinates.
Every picture of a charged species shows its charge, as a bracket with the charge or as a single charged atom that carries the whole of it; in 26 pictures the mark was missing, mostly because the counter-ion that implied it had been erased, and was added in the picture's own style, and in three main-group pictures drawn from the coordinates the drawn charges were corrected to sum to the total.
Every one of the 35 open-shell species carries its spin, as ``$S = \ldots$'', under the structure (\Cref{tab:charge-spin}).
Each edit is recorded with its entry in the released reference table.

\subsection{Dataset classification and complexity descriptors}
\label{app:classification}
\textbf{Reference-geometry annotations.}
Subgroup analyses use the fixed annotations in the released dataset; verification uses the curated reference XYZ for each entry.
For these annotations, neighbours are assigned using interatomic distances relative to the sum of covalent radii~\citep{cordero2008radii}, and every heavy atom with at least two neighbours is classified from the angles between its coordination sites, an $\eta$-bound ligand counting as one site however many of its atoms touch the metal.
In contrast, prediction--reference connectivity and coordination-site verification use xTB WBO matrices and MAPPA correspondence, as described in \Cref{app:evaluation}; the geometric neighbour rule used for these dataset annotations does not replace the WBO criteria used in verification.
For bookkeeping, the released geometry table also assigns a representative centre to each molecule: its metal if it has one, otherwise a hypervalent atom, otherwise its most coordinated non-carbon.
This automatic representative-centre label is not used to describe natural-product complexity: those molecules are analysed by size and, for metal-free entries, $n$SPS.
The Results use local-geometry classes for the VSEPR-curated subset and the defining metal's coordination number for literature complexes.
Two-coordinate centres are linear or bent by their angle; a three-coordinate centre with one near-linear angle is T-shaped, and is otherwise trigonal planar or trigonal pyramidal by the sum of its three angles; four-coordinate centres are square planar, seesaw, or tetrahedral by $\tau_4$~\citep{yang2007tau4}; and five-coordinate centres are square pyramidal or trigonal bipyramidal by $\tau_5$~\citep{addison1984tau5}.
Over the defining centres that carry no $\eta$-bound ligand, every one of these classes occupies a range of its measure separated from its neighbours by a clear margin (\Cref{tab:geomeasures}): a largest angle of at least 170\textdegree{} for linear centres against at most 151\textdegree{} for bent ones; at least 155\textdegree{} for T-shaped centres against at most 134\textdegree{} for the other three-coordinate classes; an angle sum of at least 356\textdegree{} for trigonal planar centres against at most 348\textdegree{} for trigonal pyramidal ones; $\tau_4$ below 0.2 for square planar, from 0.23 to 0.62 for seesaw, and above 0.77 for tetrahedral centres; and $\tau_5$ below 0.43 for square pyramidal and above 0.64 for trigonal bipyramidal ones.
The six- to eight-coordinate classes---octahedral, trigonal prismatic, capped octahedral, pentagonal bipyramidal, and square antiprismatic---are carried by 47 defining centres.

\begin{table}[t]
  \centering
  \caption{\textbf{How the geometry classes separate at the defining centres.} For each class of two- to five-coordinate centres, the range of the angular measure that decides it, over the 159 defining centres of these classes that carry no $\eta$-bound ligand; 85 with an $\eta$-bound ligand and 9 whose coordination number depends on how a long contact is counted are left out. Angles in degrees.}
  \label{tab:geomeasures}
  \small
  \renewcommand{\arraystretch}{1.15}
  \begin{tabular}{@{}lrlrr@{}}
  \toprule
  Class & $n$ & Measure & Range & Angle sum \\
  \midrule
  Linear & 9 & largest angle & 170.5--180.0 &  \\
  Bent & 38 & largest angle & 103.8--150.9 &  \\
  T-shaped & 6 & largest angle & 155.3--176.3 & 334.7--360.0 \\
  Trigonal planar & 17 & largest angle & 120.1--134.1 & 356.0--360.0 \\
  Trigonal pyramidal & 23 & largest angle & 87.9--120.0 & 260.0--348.0 \\
  Tetrahedral & 19 & $\tau_4$ & 0.776--0.983 &  \\
  Seesaw & 6 & $\tau_4$ & 0.228--0.617 &  \\
  Square planar & 8 & $\tau_4$ & 0.000--0.199 &  \\
  Square pyramidal & 20 & $\tau_5$ & 0.000--0.427 &  \\
  Trigonal bipyramidal & 13 & $\tau_5$ & 0.645--1.005 &  \\
  \bottomrule
  \end{tabular}
\end{table}

The classification was checked against the independent VSEPR label carried by the 59 VSEPR-curated entries and agrees for 58.
The exception is a cobalt adduct of an iodosylarene~\citep{hill2018}, selected for the T-shaped geometry of its iodine; the defining-centre rule ranks a metal above a hypervalent atom, so the classified centre is the cobalt.
The $\eta$ rule is what puts metallocenes on the same footing as other complexes: in the bis(cyclopentadienyl)molybdenum diiodide of MOR41, twelve atoms are bound to the molybdenum---the ten ring carbons and the two iodides---but each ring counts as one site, so the centre is four-coordinate and classified as tetrahedral, the bent-metallocene arrangement a chemist would name.

\textbf{Spacial score as a complexity index.}
SPS is a topological index of organic molecular complexity, and $n$SPS is its size-normalised form~\citep{krzyzanowski2023sps}.
Its use is not limited to natural products: the post-hoc complexity analysis here includes metal-free organic structures from both the natural-product and literature sources, while excluding metal complexes.
For the metal-free organic subset, SPS sums $h \cdot s \cdot r \cdot n^2$ over the heavy atoms, with $h = 3$, 2, or 1 for an sp$^3$, sp$^2$, or sp atom and 4 for other hybridisation states; $s = 2$ for a stereogenic atom or one on a stereogenic double bond and 1 otherwise; $r = 2$ for an atom in a non-aromatic ring and 1 otherwise; and $n$ the number of heavy-atom neighbours~\citep{krzyzanowski2023sps}.
$n$SPS divides SPS by the number of heavy atoms.
It is computed with RDKit~\citep{rdkit}, and was checked by hand against reference values: benzene scores 48, cyclohexane 144, and ethane 6.

\textbf{Size and complexity strata.}
Size is recorded as both total atom count, including hydrogens, and heavy-atom count from the reference XYZ files.
The heavy-atom quintiles in \Cref{tab:axes} characterise the dataset; the Results analysis uses total-atom quintiles, 18--51, 52--66, 67--80, 81--107, and 108--202, containing 59, 58, 64, 57, and 62 molecules (\Cref{tab:atom-difficulty}).
The organic complexity analysis combines 69 metal-free natural products and 15 metal-free literature structures.
The magnesium-containing chlorophyll~$a$ entry is excluded from this metal-free subset.
Joint $n$SPS quartiles have 21 molecules each, with boundaries listed in \Cref{tab:organic-difficulty}.
Literature coordination analysis uses the 156 metal-containing entries and the CN of each defining metal centre.
CN counts coordination sites, grouping an $\eta$-bound ligand as one site; a chelating ligand can occupy several sites, so CN is not the number of independent ligands.
Geometry characterisation covers all 11{,}271 centres, while the VSEPR difficulty analysis uses only the 59 entries selected for local geometry.

\subsection{Validation}
\label{app:validation}
Validation checks that each input drawing and reference XYZ represent the same chemical species, with a consistent atom inventory, connectivity, depicted stereochemistry, net charge, and spin assignment.
These checks concern the correctness of the drawing--reference pair; they do not require a common energy-minimisation method or a unique conformation.
All 300 retained pairs underwent curation checks, with source-specific procedures as follows.
VSEPR-curated depictions and any redrawn literature depictions come from their reference coordinates (\Cref{app:depictions}).
Literature depictions are matched to the corresponding deposited coordinates using the species labels in each paper; organometallic figure pairs were additionally checked atom by atom (\Cref{tab:papers}).
Each SynArchive drawing is paired with a reference for the depicted compound. The chlorophyll~$a$ reference is the ideal geometry of \href{https://www.rcsb.org/ligand/CLA}{wwPDB component CLA}.
Candidates with incomplete coordinates or a mismatched species or configuration were excluded during selection (\Cref{app:construction}).
The electron count of every reference was checked against its recorded net charge and spin multiplicity, and every picture against both (\Cref{app:depictions}); the values the verifier reads from the reference files (\Cref{app:evaluation}) agree with the descriptor table for all 300.

\subsection{Evaluation procedure}
\label{app:evaluation}

\textbf{Design.}
A prediction is not optimised before scoring, so a poor geometry cannot be relaxed into a passing one.
The released XYZ coordinates define the reference geometry for each drawing.
The electronic-structure-derived WBO matrices supply the connectivity inputs to MAPPA without converting either structure to SMILES or imposing an organic valence model. Separate local-site rules are used for the coordination-geometry check below.
MAPPA grows matched fragments outward from seed atoms, accepting an image for a reference atom only if its element agrees and every bond already matched to that atom is also present in the prediction; it branches where a fragment admits several placements, and keeps the placements related by molecular symmetry as permutation groups rather than enumerating them.
The rules below implement the molecular-identity, stereochemistry, and ligand-geometry checks described in \Cref{sec:task}.

The verifier reads the XYZ file the system saved and checks that the atom count declared on its first line matches the number of coordinate lines; a file that fails this check, or that cannot be parsed at all, is scored as unparseable output.
A prediction whose element composition differs from the reference is scored as a wrong molecule without further computation.

\textbf{Bond orders.}
Otherwise a GFN2-xTB single point~\citep{bannwarth2019gfn2} is run with xtb~6.7.1~\citep{bannwarth2021xtb} on the prediction and on the reference, both at the net charge and spin multiplicity recorded on the reference's comment line---for example, \texttt{q = 1, multiplicity = 2} or \texttt{q = 1, S = 0.5}, where the multiplicity is $2S+1$---or as a neutral singlet where the comment records none, and the Wiberg bond-order matrix~\citep{wiberg1968} of each is read from the output.
A WBO floor of 0.15 defines the edges of the MAPPA input graphs. This graph-growth cutoff is distinct from the pair-specific WBO-change thresholds below and from the coordination-site retention rule.

\textbf{Atom-to-atom mapping.}
The two WBO graphs are aligned by atom-to-atom mapping with MAPPA~\citep{zou2026mappa}, using three seed orderings, a matching tolerance of 1, and a graph floor of 0.15.
The branch cap is 10{,}000 and cut sweep is disabled.
The verifier can search in either mapping direction. A successful MAPPA search supplies a complete bijection $m$ between reference and prediction atoms, hydrogens included, either directly or by realising a zero-event candidate from a retained native mapping family.
For every pair of reference atoms $i,j$ with bond orders $W_R(i,j)$ and $W_P(m(i),m(j))$, the pair is \emph{broken} when $W_R - W_P > \delta$ and \emph{formed} when $W_P - W_R > \delta$, with $\delta=0.50$ for pairs containing a metal or another MAPPA fractional-bond element (including B, Si, P, S and Cl), and $\delta=0.75$ for other pairs.
The selected correspondence must have zero broken and zero formed bonds. Predictions scored as mismatched molecular identity fail this gate.

\textbf{Stereochemistry.}
The reference defines the stereochemical constraints, using a covalent subgraph that requires WBO at least 0.4 and an interatomic distance below 1.35 times the sum of the covalent radii.
Element-initialised Weisfeiler--Lehman colours~\citep{weisfeiler1968} distinguish substituents: four distinguishable neighbours define a tetrahedral frame, and three distinguishable neighbours at P, As, S, or Se can define a pyramidal one. Metals are checked separately by their ligand sites.
For tetrahedral frames, the signed volume of the four neighbours is normalised by the cube of their geometric-mean pair distance; for pyramidal frames the measure is the determinant of the three unit bond vectors.
A reference tetrahedral frame whose normalised magnitude is at most 0.02 defines no configuration and is recorded as such. Once a configuration is defined, a predicted measure with magnitude at most 0.02 is a loss of configuration and fails; a change of sign is an inversion and also fails.
For E/Z geometry, covalent edges with WBO at least 1.5 define the candidate double bonds, with unannotated rings of seven atoms or fewer excluded. Projections of distinguishable end substituents onto the plane normal to the bond axis give their relative side; the analogous signed cross product checks supported cumulene axes.
For these bond and axis checks, substituent projections of length at most 0.10~\AA{} are degenerate. After normalisation, a side or axial-orientation measure with magnitude at most 0.10 is also degenerate. A configuration defined in the reference but degenerate in the prediction fails.
The verifier also supports explicitly annotated single-bond stereogenic axes; no such annotations are supplied for the present 300 entries, so unrestricted single-bond atropisomerism is not claimed as covered.
Metal contacts are excluded from inferred tetrahedral carbon and nitrogen frames. A guanidine C=N bond requires an explicit E/Z annotation, and a tetrahedral centre marked unspecified in the drawing imposes no handedness constraint. A stereochemical failure triggers a search over symmetry-equivalent assignments of the reference atoms used in the stereo checks. An alternative assignment is accepted only if it has zero bond events and passes both coordinate checks. A ligand-geometry failure alone does not trigger this search.

\textbf{Ligand geometry.}
The coordination check uses local binding sites defined on the reference structure~\citep{sobez2020molassembler} and compares their arrangement under the selected MAPPA correspondence.
For a metal or supported hypervalent centre $M$, a donor $D$ enters the reference frame when $W_R(M,D)\geq 0.15$. An angle comparison requires at least two distinct sites.
Coordination-site definition and retention impose no absolute, radius-based or reference-relative bond-length cutoff; coordinates supply the site directions and angles.
At metal centres, adjacent non-hydrogen, nonmetal donors connected in a separate WBO-defined covalent subgraph (floor 0.4) by a mutual WBO of at least 1.1 form one haptic site; its direction is represented by the donor centroid, following local-site descriptions of $\pi$-bound ligands~\citep{cirera2005shape}.
Metal bridges and ordinary single bonds between donors do not by themselves merge separate positions.

At a metal centre, a site is lost when every eligible donor has $W_R-W_P>0.50$. At a supported nonmetal centre, site retention requires at least one donor with $W_P\geq0.10$.
For a single occupied donor position, a previously unoccupied terminal atom of the same element attached to the same nonmetal neighbour can preserve the site, as with the two oxygens of a sulfonyl group.
The donor is selected by WBO, retaining the mapped reference donor on an exact tie, without using distance, fitting angles or changing the MAPPA correspondence; two occupied equivalent donors remain two required positions.
A lost site or collapsed centre-to-site direction fails.
Additional bonding is assessed by the MAPPA WBO-change rule; crossing the graph floor alone does not create an independent failure from a new prediction coordination count.

Centre-to-site directions define the inter-site angles.
Haptic grouping, equivalent-donor identification and donor types use the WBO-defined covalent subgraph, independently of distance; donor types are assigned after removing the checked centre, so centre--donor WBO differences do not split equivalent ligands.
Angles are compared as sorted profiles within the same chemical site-pair class, distinguishing chelate bite angles from angles between separate ligands.
For the resulting $K$ corresponding angle values, $\Delta\theta_k=\theta_{P,k}-\theta_{R,k}$, acceptance requires
\[
\sqrt{\frac{1}{K}\sum_{k=1}^{K}(\Delta\theta_k)^2}\leq20^\circ,
\qquad
\max_k|\Delta\theta_k|\leq60^\circ.
\]
Exceeding either limit fails, including a value just above a boundary; the same limits apply to all checked centres.
These are operational tolerances for gross rearrangements, not fine conformational-accuracy criteria or universal chemical boundaries.
The angle profiles do not provide a complete test of metal-centred chirality, chelate helicity or arbitrary changes in hapticity.

\textbf{Outcome accounting.}
A submission is flagged if its element-labelled Cartesian coordinates match the reference after rounding to four decimal places in \AA{}, irrespective of atom order. This flag does not change the score; none of the submitted structures from the 2{,}700 evaluated attempts carries it.
A failed single point on the prediction is recorded as a failed bond-order calculation. There are four such cases among the 2{,}700 evaluated attempts.
Every attempt has a pass or fail outcome. Scoring uses a 300-second limit, with one 900-second retry for timed-out cases. A remaining timeout, a MAPPA search that does not establish the required correspondence, or a failed bond-order calculation is scored as failure. Every case remains in the denominator: overall accuracy is passes divided by 300, and a source or classification-axis rate is passes divided by all entries of that group.
Each molecule is scored on a single attempt (\Cref{sec:agentsetup}), so no aggregation over samples is needed.

\clearpage
\subsection{Dataset characterisation}
\label{app:characterisation}

\Cref{tab:stats} summarises what each source contributes, \Cref{tab:axes} gives the three classification axes by source, \Cref{tab:geomdist} the geometry class at the defining centre for each source, and \Cref{tab:centres} the class of every one of the 11{,}271 classified centres; \Cref{tab:elements} lists the 52 elements and how many molecules contain each, and \Cref{tab:charge-spin} the net charges and spin multiplicities.

\begin{table}[H]
  \centering
  \caption{\textbf{Composition of \bench{}.} The three sources and what each contributes. $n$SPS is the size-normalised spacial score~\citep{krzyzanowski2023sps}, given for the 84 metal-free natural-product and literature structures used in the organic analysis (\Cref{app:classification}); the ``Geometry classes'' entry counts the distinct VSEPR classes realised at the defining centres of that source.}
  \label{tab:stats}
  \small
  \renewcommand{\arraystretch}{1.25}
  \begin{tabular}{@{}lrrrr@{}}
  \toprule
   & VSEPR-curated & Natural prod. & Literature & All \\
  \midrule
  Molecules & 59 & 70 & 171 & 300 \\
  Atoms (median) & 65 & 54.5 & 90 & 74 \\
  \midrule
  Heavy atoms (median) & 35 & 28 & 46 & 39 \\
  Heavy atoms (range) & 12--79 & 16--65 & 8--106 & 8--106 \\
  $n$SPS (median) & -- & 43.4 & 44.9 & 43.9 \\
  Rings (median) & 3 & 5 & 5 & 5 \\
  \midrule
  Contains a metal (\%) & 75 & 1 & 91 & 67 \\
  Charged (\%) & 12 & 1 & 42 & 27 \\
  Geometry classes & 12 & 4 & 15 & 15 \\
  Distinct elements & 40 & 9 & 44 & 52 \\
  \bottomrule
  \end{tabular}
\end{table}

\begin{table}[H]
  \centering
  \caption{\textbf{The three classification axes of \bench{}, by source.} Size bins are the quintiles of heavy-atom count over all 300 molecules. Complexity bins are a post-hoc description using $n$SPS as an organic complexity index, with joint quartiles over 69 metal-free natural products and 15 metal-free literature structures. They do not stratify metal-complex difficulty and are separate from candidate selection (\Cref{app:construction,app:classification}).}
  \label{tab:axes}
  \small
  \renewcommand{\arraystretch}{1.25}
  \begin{tabular}{@{}lrrrr@{}}
  \toprule
   & VSEPR-curated & Natural prod. & Literature & All \\
  \midrule
  \textbf{Size} \scriptsize(heavy atoms) &  &  &  &  \\
  \quad 8--24 & 12 & 24 & 21 & 57 \\
  \quad 25--33 & 17 & 27 & 18 & 62 \\
  \quad 34--43 & 16 & 10 & 35 & 61 \\
  \quad 44--51 & 9 & 5 & 46 & 60 \\
  \quad 52--106 & 5 & 4 & 51 & 60 \\
  \midrule
  \textbf{Organic complexity} \scriptsize($n$SPS) &  &  &  &  \\
  \quad 18.00--33.76 & -- & 18 & 3 & 21 \\
  \quad 34.00--43.86 & -- & 18 & 3 & 21 \\
  \quad 43.88--52.27 & -- & 17 & 4 & 21 \\
  \quad 52.67--108.00 & -- & 16 & 5 & 21 \\
  \midrule
  \textbf{Centre coordination} &  &  &  &  \\
  \quad 2 & 2 & 29 & 23 & 54 \\
  \quad 3 & 12 & 34 & 22 & 68 \\
  \quad 4 & 13 & 7 & 60 & 80 \\
  \quad 5 & 16 & 0 & 35 & 51 \\
  \quad 6 & 12 & 0 & 21 & 33 \\
  \quad 7 & 4 & 0 & 6 & 10 \\
  \quad 8 & 0 & 0 & 4 & 4 \\
  \bottomrule
  \end{tabular}
\end{table}

\begin{table}[H]
  \centering
  \caption{\textbf{Geometry class at the defining centre, by source.} CN is the coordination number the class implies. Every heavy atom with at least two neighbours is classified; this table reports the single defining centre each molecule is additionally assigned (\Cref{app:classification}).}
  \label{tab:geomdist}
  \small
  \renewcommand{\arraystretch}{1.25}
  \begin{tabular}{@{}lcrrrr@{}}
  \toprule
  Geometry class & CN & VSEPR-curated & Natural prod. & Literature & All \\
  \midrule
  Bent & 2 & 1 & 29 & 11 & 41 \\
  Linear & 2 & 1 & 0 & 12 & 13 \\
  T-shaped & 3 & 4 & 0 & 4 & 8 \\
  Trigonal planar & 3 & 4 & 15 & 16 & 35 \\
  Trigonal pyramidal & 3 & 4 & 19 & 2 & 25 \\
  Seesaw & 4 & 4 & 0 & 36 & 40 \\
  Square planar & 4 & 7 & 0 & 9 & 16 \\
  Tetrahedral & 4 & 2 & 7 & 15 & 24 \\
  Square pyramidal & 5 & 11 & 0 & 27 & 38 \\
  Trigonal bipyramidal & 5 & 5 & 0 & 8 & 13 \\
  Octahedral & 6 & 12 & 0 & 20 & 32 \\
  Trigonal prismatic & 6 & 0 & 0 & 1 & 1 \\
  Capped octahedral & 7 & 0 & 0 & 4 & 4 \\
  Pentagonal bipyramidal & 7 & 4 & 0 & 2 & 6 \\
  Square antiprismatic & 8 & 0 & 0 & 4 & 4 \\
  \bottomrule
  \end{tabular}
\end{table}

\begin{table}[H]
  \centering
  \caption{\textbf{Geometry class over every classified centre, by source.} All 11{,}271 heavy-atom centres of the benchmark, classified as in \Cref{app:classification}; \Cref{tab:geomdist} reports the defining centres alone.}
  \label{tab:centres}
  \small
  \renewcommand{\arraystretch}{1.25}
  \begin{tabular}{@{}lrrrr@{}}
  \toprule
  Geometry class & VSEPR-curated & Natural prod. & Literature & All \\
  \midrule
  Trigonal planar & 1133 & 617 & 3457 & 5207 \\
  Tetrahedral & 575 & 1072 & 3191 & 4838 \\
  Bent & 132 & 226 & 214 & 572 \\
  Seesaw & 5 & 1 & 202 & 208 \\
  Trigonal pyramidal & 6 & 88 & 32 & 126 \\
  Linear & 23 & 2 & 94 & 119 \\
  Square pyramidal & 11 & 0 & 53 & 64 \\
  T-shaped & 6 & 0 & 32 & 38 \\
  Octahedral & 12 & 0 & 24 & 36 \\
  Square planar & 7 & 0 & 15 & 22 \\
  Trigonal bipyramidal & 5 & 0 & 16 & 21 \\
  Pentagonal bipyramidal & 4 & 0 & 2 & 6 \\
  Trigonal prismatic & 0 & 0 & 6 & 6 \\
  Capped octahedral & 0 & 0 & 4 & 4 \\
  Square antiprismatic & 0 & 0 & 4 & 4 \\
  \midrule
  All centres & 1919 & 2006 & 7346 & 11271 \\
  \bottomrule
  \end{tabular}
\end{table}

\begin{table}[H]
  \centering
  \caption{\textbf{Element coverage of \bench{}.} The 52 elements of the benchmark and, for each, the number of the 300 molecules that contain it, read from the molecular formulas.}
  \label{tab:elements}
  \small
  \renewcommand{\arraystretch}{1.25}
  \begin{tabular}{@{}l>{\raggedright\arraybackslash}p{10.5cm}@{}}
  \toprule
   & Element and number of molecules containing it \\
  \midrule
  Non-metals and metalloids (16) & C~300, H~299, N~200, O~179, P~80, Cl~49, Si~35, F~33, S~30, B~13, Br~10, Ge~8, I~7, Se~3, Sb~2, Te~1 \\
  Main-group metals (9) & Ga~6, Sn~5, K~3, Al~2, In~2, Ca~1, Mg~1, Na~1, Tl~1 \\
  Transition metals (27) & Ir~30, Ni~22, Rh~22, Ru~19, Pd~15, Co~13, Au~11, Cu~10, Cr~8, Ti~8, Zn~8, Zr~7, Os~5, W~5, Fe~4, Mo~4, Y~4, Ag~3, Cd~3, Tc~3, V~3, Hf~2, Hg~1, Nb~1, Pt~1, Re~1, Ta~1 \\
  \bottomrule
  \end{tabular}
\end{table}

\begin{table}[H]
  \centering
  \caption{\textbf{Net charge and spin multiplicity, by source.} As recorded for each reference and used for its xTB single point; a charged species shows its charge in the depiction, and an open-shell one its spin (\Cref{app:depictions}).}
  \label{tab:charge-spin}
  \small
  \renewcommand{\arraystretch}{1.25}
  \begin{tabular}{@{}lrrrr@{}}
  \toprule
   & VSEPR-curated & Natural prod. & Literature & All \\
  \midrule
  \textbf{Net charge} &  &  &  &  \\
  \quad $-2$ & 0 & 0 & 1 & 1 \\
  \quad $-1$ & 3 & 0 & 9 & 12 \\
  \quad 0 & 52 & 69 & 99 & 220 \\
  \quad $+1$ & 4 & 1 & 50 & 55 \\
  \quad $+2$ & 0 & 0 & 11 & 11 \\
  \quad $+3$ & 0 & 0 & 1 & 1 \\
  \midrule
  \textbf{Spin multiplicity} &  &  &  &  \\
  \quad 1 & 58 & 70 & 137 & 265 \\
  \quad 2 & 1 & 0 & 17 & 18 \\
  \quad 3 & 0 & 0 & 8 & 8 \\
  \quad 4 & 0 & 0 & 4 & 4 \\
  \quad 5 & 0 & 0 & 2 & 2 \\
  \quad 6 & 0 & 0 & 1 & 1 \\
  \quad 9 & 0 & 0 & 2 & 2 \\
  \bottomrule
  \end{tabular}
\end{table}

\FloatBarrier
\subsection{Data availability and reproduction}
\label{app:availability}
The benchmark depictions, reference coordinates, submitted XYZ structures and verifier are available at \url{https://github.com/the-matter-lab/MolDraft}. The archived dataset is also available on Zenodo at \url{https://doi.org/10.5281/zenodo.24511623}.
The \href{https://github.com/the-matter-lab/MolDraft/tree/d4499ae0a6ae54229b799ee283622ea469d4f00c/results}{session archive} contains the main and ablation transcripts. The result tables link every attempt to its model, molecule identifier, submitted answer where present, and binary verification verdict, including all 2,700 main-experiment attempts and the 300 two-tool ablation attempts. The paper results are in \texttt{results/latest/paper}. Results for the three additional models are in \texttt{results/latest/additional}.
Entries whose coordinates come from PubChem, PubChemQC, the wwPDB Chemical Component Dictionary, the Crystallography Open Database, or the Supporting Information and data sets of publications are distributed with their coordinates.
The 41 transition-metal entries use tmQM reference geometries derived from CSD structures and are identified by their CSD refcodes.
Third-party image files retain their source-specific copyright and licence conditions (\Cref{app:sources}).
Their accompanying source records identify the original paper or webpage, the figure or species label, the input filename, and the recorded edits.

\section{Evaluation setup in detail}
\label{app:setup}
This section provides in full the agent, the input condition, the prompt, the LLMs, and the environment and budget summarised in \Cref{sec:setup}.

\bench{} can be attempted by anything that reads an image and writes a file.
We evaluate one agentic system---a multimodal language model with four tools, which plans, executes code, inspects what it has produced, and revises---and run it in turn on nine LLMs (\Cref{sec:models}).
Holding the scaffold fixed while the model varies is the design choice this section argues for.
Reading the drawing and building the structure are separable failures, and the analysis of \Cref{sec:results} separates them.

\subsection{Four tools}

The system is a single language-model node with four tools and no tool that builds a structure.
\texttt{view\_png\_file} returns an image file to the model, which is how it reads the drawing; \texttt{bash\_kernel} runs a shell command in the working directory, so the agent can write and execute code with whatever libraries the environment provides; \texttt{view\_molecule} renders a coordinate file with atom indices and element labels and lets the agent rotate the view, so it can inspect what it has built; and \texttt{constraint\_xtb} runs a GFN2-xTB geometry optimisation~\citep{bannwarth2019gfn2} on a coordinate file, optionally under distance, angle, or dihedral constraints, so a structure assembled by hand can be relaxed without leaving the sandbox.
Everything else---parsing the drawing, choosing a strategy, building the atom inventory, placing the atoms---the agent must do itself.

\textbf{Relation to El Agente Estructural.}
The system is a reduced configuration of El Agente Estructural~\citep{choi2026estructural}, an agent for building and editing molecular geometries that is itself built on the El Agente framework~\citep{zou2025elagente}.
It keeps Estructural's agent loop and four of its tools and removes everything else: the domain-specific operators that build and edit geometries, delegation to other agents, and semantic memory.
Its system prompt is replaced by the task description reproduced in \Cref{app:prompt}.

\textbf{Why the scaffolding is thin.}
With no tool that builds a structure, the agent's result is close to what its LLM can plan and code, and the comparison between LLMs is a comparison of models rather than of tooling.
What a richer scaffold would add is a separate question, and one that \bench{} can answer by the same protocol.
The budget under which the agent runs is given in \Cref{sec:agentsetup}.

\textbf{Two-tool ablation with gpt-6.1-sol.}
The reduced configuration exposes only \texttt{view\_png\_file} and \texttt{bash\_kernel}, retaining the four-tool configuration's model, high reasoning effort, prompts, 300 molecules, 100-call budget, sandbox, and verification procedure.
Direct viewing of the input drawing is preserved, and the shell remains available for coding and installed software; only the dedicated \texttt{view\_molecule} and \texttt{constraint\_xtb} interfaces are removed.
The comparison is designed to assess the difference between basic image viewing and shell execution and the four-tool scaffold.
Results and resource use are compared in \Cref{app:ablation}.

\subsection{One scaffold, nine LLMs}

We compare nine multimodal LLMs on the same chemical task using a shared agent, input format, and evaluation protocol.

The comparison is meaningful to the extent that everything but the LLM is held fixed.
The \emph{scaffold} is identical: the same single node, the same four tools, and the same chemistry instructions (\Cref{app:prompt}).
The \emph{execution environment} is identical, so a difference in outcome cannot be attributed to installed software.
The \emph{budget} is identical in the number of model calls, because an agent given more steps is a different system from the same agent given fewer.
Two settings differ by necessity rather than by choice.
Providers expose reasoning through different controls, and we set each model to the same nominal effort wherever its interface allows (\Cref{tab:models}); and two LLMs carry an additional per-session token cap (\Cref{app:execution-budget}).
A difference in resolve rate between two LLMs is therefore a difference between the models as they use these tools.

\textbf{Protocol.}
The scaffold is the authors' own.
The evaluation protocol is the same for every system, and every answer is scored by the same verifier (\Cref{app:evaluation}); no system was tuned on \bench{}; and the tool-invocation logs for all systems are released with the benchmark.

\subsection{Input condition}

A skeletal drawing is a compressed representation.
Hydrogens on carbon are not drawn, and groups are written as condensed labels such as \ce{PMe2} or \ce{SiH3}, so a system must expand the drawing into a complete atom inventory before any geometry can be built.

\begin{table}[H]
  \centering
  \caption{\textbf{Atom inventory recovered from the depiction alone.} \emph{Returned:} structures saved and parsed, of 300. \emph{Correct composition:} returned structures whose element composition matches the reference, irrespective of connectivity or geometry. \emph{Share of failures:} the fraction of each LLM's failed attempts that fail on composition, before connectivity is examined.}
  \label{tab:composition}
  \small
  \renewcommand{\arraystretch}{1.25}
  \begin{tabular}{@{}lrrr@{}}
  \toprule
  Model & Returned & \makecell{Correct\\composition (\%)} & \makecell{Share of\\failures (\%)} \\
  \midrule
  \texttt{gpt-6-astra} & 300 & 97.0 & 20 \\
  \texttt{gpt-6.1-sol} & 300 & 95.3 & 24 \\
  \texttt{claude-opus-5} & 275 & 90.9 & 24 \\
  \texttt{claude-opus-5-5} & 300 & 90.3 & 27 \\
  \texttt{claude-fable-5-1} & 292 & 88.0 & 29 \\
  \texttt{qwen3.8-max} & 257 & 79.8 & 35 \\
  \texttt{gpt-6-sol} & 300 & 74.0 & 55 \\
  \texttt{kimi-k3} & 282 & 71.3 & 52 \\
  \texttt{gpt-6-luna} & 299 & 54.5 & 62 \\
  \bottomrule
  \end{tabular}
\end{table}

\textbf{Depiction only.}
We evaluate a single input condition: the image alone, as a chemist would encounter it.
No molecular formula, atom count, or atom list accompanies it, so expanding the drawing into the right set of atoms is part of the task, and a structure with the wrong atoms fails as a wrong molecule.

\textbf{Atom bookkeeping, measured.}
Prior work reports that frontier models emitting coordinates from a string input return the correct number of atoms in a small minority of cases, which places a ceiling on any geometric assessment made in that setting~\citep{cavanagh2026geometry}.
\Cref{tab:composition} reports, for each LLM, the fraction of returned structures whose element composition matches the reference, independently of whether connectivity or geometry is correct.
It ranges from 54.5\% to 97.0\%.
For the strongest LLMs bookkeeping is a minor share of failure---a wrong composition accounts for 20\% of the failed attempts of gpt-6-astra and 24\% of those of gpt-6.1-sol---while for the weakest it is the largest, 62\% for gpt-6-luna.
Conditions that hand the system the formula or the full atom list would separate bookkeeping from geometry; we leave them to future work (\Cref{sec:discussion}).

\subsection{Prompt format}

The agent receives a short system prompt, which describes its role and the requirements on the output file, and a user message that names the PNG file in its working directory and asks for the structure.
The prompt states no scoring criterion, shows no example XYZ block, names no atom, prescribes no method---it neither suggests nor forbids SMILES---and does not mention that network access is unavailable.
The user prompt is given verbatim in \Cref{sec:task}, and the system prompt in \Cref{app:prompt}.

\subsection{Language models}

The agent is run with nine multimodal language models from four developers (\Cref{tab:models}): four from OpenAI, three from Anthropic, and one each from Alibaba and Moonshot AI.
Because the input is an image, only multimodal models can attempt \bench{} at all; this excludes text-only models that appear in other chemistry benchmarks, and is itself a constraint worth stating.
Each model is called by its exact API identifier, with the same agent, prompt, and environment.
Reasoning is enabled for every model at a nominal effort of high, except qwen3.8-max, which is run at medium.
Tool calls are issued one at a time, and each molecule is attempted once, except where the provider failed (\Cref{sec:agentsetup}).

\subsection{Execution environment and budget}
\label{app:execution-budget}

\textbf{Execution environment.}
Every run uses the same sandbox.
Each shell command the agent issues runs inside an unprivileged user, mount, and network namespace created for that command.
The namespace's only network interface is loopback, so no command can reach the internet, and the directories that hold the reference geometries, the private index that maps anonymised file names to molecules, and the verifier's cached calculations are covered by empty file systems, so from inside the namespace they do not exist.
For the duration of a run the same directories are also moved out of the shared file tree, and the proxy variables of every command are pointed at a closed port.
The working directory of a session holds the anonymised PNG and whatever the agent writes; nothing else from the benchmark is reachable.
Holding the environment fixed means that a difference in outcome between two LLMs cannot be attributed to a difference in installed software.
The same execution environment and tool interfaces are used for all LLMs.

\textbf{Budget.}
A session is limited to 100 model calls.
For kimi-k3 and qwen3.8-max it is additionally limited to five million tokens, input and output summed over every call, since every call resends the whole conversation and a turn limit alone does not bound what a session consumes.
One session per model and molecule is scored.
A session that reaches a limit ends without a structure and is scored as no answer.

\textbf{Provider failures.}
A session can also end because the model's provider returns an error, which says nothing about the system.
Of the 2{,}700 sessions, 26 of qwen3.8-max's ended this way, 13 on an upstream error and 13 on an empty reply; they were run once more, and the second attempt is the one scored.
Three sessions of gpt-6-luna were cut off by its provider mid-run and were also run once more; all three remained unsuccessful on the second attempt and are included as failures in the 300-attempt denominator.

\textbf{Instrumentation.}
For every run we record the tools invoked and the code executed, so that the strategy a system adopted---and whether it detected that the strategy had failed---can be reported alongside whether it succeeded.

\textbf{Coverage.}
All nine LLMs are run on the full benchmark under the depiction-only condition, one scored session per molecule.

\FloatBarrier
\section{System prompt}
\label{app:prompt}
The shared chemistry instructions in the system prompt, reproduced as used in the archived runs:
\begin{quote}\small\ttfamily\raggedright
You are a helpful chemistry assistant. You build 3D molecular structures from the chemical information you are given, including 2D drawings supplied as images.\\[0.5em]
When you are given a 2D drawing of a molecule, your task is to produce the 3D structure that drawing depicts and save it as `answer.xyz' in your working directory.\\[0.5em]
Requirements for `answer.xyz':\\
-- Include all explicit hydrogens.\\
-- Use a right-handed Cartesian coordinate system, units of angstroms.\\
-- Standard XYZ format: first line is the total atom count, second line a one-line comment, each remaining line `Element X Y Z'.\\[0.5em]
When you are done, confirm that `answer.xyz' exists in the working directory.
\end{quote}
The hydrogen instruction above is the wording used in all archived main and ablation runs; evaluation requires all hydrogens, including those implicit in the drawing.
For future runs, it should read: ``Include all hydrogens, including those implicit in the drawing.''

For qwen3.8-max and kimi-k3, the system message additionally included the following tool-use instruction:
\begin{quote}\small\ttfamily\raggedright
[IMPORTANT CONSTRAINT] You must NOT call multiple tools in parallel. Execute tools ONE AT A TIME, sequentially. Wait for each tool's result before calling the next tool. Never batch multiple tool calls in a single response.
\end{quote}
\FloatBarrier
\section{Full results}
\label{app:results}
\Cref{tab:results} gives every LLM's resolve rate, overall and by source, and its failures by category.

\begin{table}[H]
  \centering
  \caption{\textbf{Main results.} Resolve rate (\%) of each LLM over all 300 molecules, overall and by source, followed by the number of molecules, of 300, in each failure category: no answer, including a session that reached its budget or was cut off by the provider; a wrong element composition; other molecular-identity mismatches (connectivity); stereochemical inversion or loss of configuration; gross ligand-geometry mismatch; failed bond-order calculations; and other verification failures. All failures remain in the denominator: 300 overall, or the full source size for source-specific rates.}
  \label{tab:results}
  \small
  \renewcommand{\arraystretch}{1.25}
  \setlength{\tabcolsep}{2.3pt}
  \begin{tabular}{@{}lrrrrrrrrrrr@{}}
  \toprule
  & \multicolumn{4}{c}{Resolve rate (\%)} & \multicolumn{7}{c}{Molecules, by outcome} \\
  \cmidrule(lr){2-5}\cmidrule(l){6-12}
  Model & All & \makecell{VSEPR-\\curated} & \makecell{Natural\\prod.} & Literature & \makecell{No\\answer} & \makecell{Compo-\\sition} & \makecell{Connec-\\tivity} & \makecell{Stereo-\\chem.} & \makecell{Ligand\\geom.} & xTB & \makecell{Other\\failure} \\
  \midrule
  \texttt{gpt-6-astra} & 85.0 & 88.1 & 88.6 & 82.5 & 0 & 9 & 3 & 9 & 21 & 0 & 3 \\
  \texttt{gpt-6.1-sol} & 80.7 & 84.7 & 85.7 & 77.2 & 0 & 14 & 6 & 14 & 24 & 0 & 0 \\
  \texttt{claude-opus-5} & 65.0 & 67.8 & 84.3 & 56.1 & 25 & 25 & 5 & 8 & 39 & 0 & 3 \\
  \texttt{claude-opus-5-5} & 63.7 & 67.8 & 87.1 & 52.6 & 0 & 29 & 19 & 13 & 45 & 0 & 3 \\
  \texttt{claude-fable-5-1} & 60.0 & 67.8 & 67.1 & 54.4 & 8 & 35 & 17 & 26 & 32 & 0 & 2 \\
  \texttt{gpt-6-sol} & 52.3 & 69.5 & 55.7 & 45.0 & 0 & 78 & 7 & 26 & 27 & 0 & 5 \\
  \texttt{qwen3.8-max} & 50.0 & 76.3 & 67.1 & 33.9 & 43 & 52 & 11 & 9 & 23 & 0 & 12 \\
  \texttt{kimi-k3} & 47.7 & 57.6 & 52.9 & 42.1 & 18 & 81 & 14 & 13 & 20 & 0 & 11 \\
  \texttt{gpt-6-luna} & 26.3 & 49.2 & 34.3 & 15.2 & 1 & 136 & 19 & 18 & 18 & 4 & 25 \\
  \bottomrule
  \end{tabular}
\end{table}

\Cref{tab:efficiency} gives the nine models' passes, cost per molecule, and tokens per molecule, with all three quantities normalised by 300 attempts.
\Cref{tab:vsepr-difficulty,tab:organic-difficulty,tab:coordination-difficulty,tab:atom-difficulty} summarise the separate category and atom-count analyses; subset rates retain all selected molecules in their denominators.

\begin{table}[H]
  \centering
  \caption{\textbf{Accuracy and resources for the nine LLMs.} Passes, cost and total input plus output tokens are each divided by all 300 attempted molecules. Cached tokens are already included in input tokens.}
  \label{tab:efficiency}
  \small
  \renewcommand{\arraystretch}{1.25}
  \begin{tabular}{@{}lrrrr@{}}
  \toprule
  Model & Resolved / 300 & Rate (\%) & USD / molecule & Tokens / molecule \\
  \midrule
  \texttt{gpt-6-astra} & 255/300 & 85.0 & 0.5634 & 101,681 \\
  \texttt{gpt-6.1-sol} & 242/300 & 80.7 & 0.0771 & 81,815 \\
  \texttt{claude-opus-5} & 195/300 & 65.0 & 2.5431 & 1,136,292 \\
  \texttt{claude-opus-5-5} & 191/300 & 63.7 & 0.1641 & 73,229 \\
  \texttt{claude-fable-5-1} & 180/300 & 60.0 & 1.9657 & 162,169 \\
  \texttt{gpt-6-sol} & 157/300 & 52.3 & 0.1473 & 173,248 \\
  \texttt{qwen3.8-max} & 150/300 & 50.0 & 0.8313 & 821,971 \\
  \texttt{kimi-k3} & 143/300 & 47.7 & 2.1568 & 1,645,860 \\
  \texttt{gpt-6-luna} & 79/300 & 26.3 & 0.0153 & 373,236 \\
  \bottomrule
  \end{tabular}
\end{table}

\subsection{Within-category difficulty}
\label{app:category-difficulty}

\textbf{Geometry within the VSEPR-curated subset.}
\label{sec:res-geometry}
Among the 59 VSEPR-curated entries, octahedral structures average 81.5\% success ($n=12$), square-pyramidal structures 72.7\% ($n=11$), square-planar structures 73.0\% ($n=7$), and trigonal-bipyramidal structures 55.6\% ($n=5$; \Cref{tab:vsepr-difficulty}).
The other classes contain only one to four molecules each, limiting comparisons between geometries.
These are reconstruction rates for molecules classified by their reference centres; the verifier checks coarse ligand arrangements but does not quantify fine bond-angle accuracy or independently classify every predicted centre.

\textbf{Complexity of organic natural products and literature molecules.}
\label{sec:res-complexity}
The joint metal-free organic subset contains 69 natural products and 15 literature molecules.
Across four equal-sized $n$SPS bins ($n=21$ each), mean success rates are 70.4\%, 72.5\%, 74.6\%, and 48.1\% from lowest to highest complexity (\Cref{tab:organic-difficulty}).
The highest-complexity quartile is the hardest, while the first three quartiles are not ordered by increasing difficulty.
Thus, high organic structural complexity is associated with reduced success, but $n$SPS alone does not determine the outcome.

\textbf{Coordination number in literature metal complexes.}
\label{app:coordination-difficulty}
The 156 metal-containing literature structures are grouped by the coordination number (CN) of their defining metal centre, rather than by the number of independent ligands.
Mean rates are 56.6\% at CN~3 ($n=22$), 56.4\% at CN~4 ($n=53$), 35.6\% at CN~5 ($n=35$), and 52.4\% at CN~6 ($n=21$; \Cref{tab:coordination-difficulty}).
CN~2 averages 50.4\%, while CN~7 and 8 average 59.3\% and 61.1\% but contain only six and four molecules.
The absence of a monotonic decline indicates that coordination number alone does not order the difficulty of these complexes.

\begin{table}[H]
  \centering
  \caption{\textbf{VSEPR-curated structures by defining-centre geometry.} Only the 59 VSEPR-curated entries are included. Rates include verifier errors in each full subset denominator. Mean and range refer to the nine LLMs.}
  \label{tab:vsepr-difficulty}
  \small
  \renewcommand{\arraystretch}{1.25}
  \begin{tabular}{@{}lrrrrr@{}}
  \toprule
  Group & $n$ & \makecell{Mean of\\nine (\%)} & Range (\%) & \makecell{gpt-6-astra\\(\%)} & \makecell{gpt-6.1-sol\\(\%)} \\
  \midrule
  bent & 1 & 77.8 & 0.0--100.0 & 100.0 & 100.0 \\
  linear & 1 & 88.9 & 0.0--100.0 & 100.0 & 100.0 \\
  T-shaped & 4 & 66.7 & 25.0--100.0 & 75.0 & 75.0 \\
  trigonal planar & 4 & 86.1 & 75.0--100.0 & 100.0 & 100.0 \\
  trigonal pyramidal & 4 & 55.6 & 25.0--100.0 & 100.0 & 75.0 \\
  seesaw & 4 & 47.2 & 25.0--75.0 & 50.0 & 50.0 \\
  square planar & 7 & 73.0 & 57.1--85.7 & 85.7 & 85.7 \\
  tetrahedral & 2 & 66.7 & 0.0--100.0 & 100.0 & 100.0 \\
  square pyramidal & 11 & 72.7 & 45.5--100.0 & 100.0 & 81.8 \\
  trigonal bipyramidal & 5 & 55.6 & 20.0--100.0 & 60.0 & 80.0 \\
  octahedral & 12 & 81.5 & 58.3--91.7 & 91.7 & 91.7 \\
  pentagonal bipyramidal & 4 & 58.3 & 0.0--100.0 & 100.0 & 100.0 \\
  \bottomrule
  \end{tabular}
\end{table}

\begin{table}[H]
  \centering
  \caption{\textbf{Organic structures by $n$SPS.} Quartiles are calculated jointly over 69 metal-free natural products and 15 metal-free literature structures. The magnesium-containing chlorophyll entry is excluded. Rates include verifier errors in each full subset denominator. Mean and range refer to the nine LLMs.}
  \label{tab:organic-difficulty}
  \small
  \renewcommand{\arraystretch}{1.25}
  \begin{tabular}{@{}lrrrrr@{}}
  \toprule
  Group & $n$ & \makecell{Mean of\\nine (\%)} & Range (\%) & \makecell{gpt-6-astra\\(\%)} & \makecell{gpt-6.1-sol\\(\%)} \\
  \midrule
  18.00--33.76 & 21 & 70.4 & 28.6--90.5 & 90.5 & 90.5 \\
  34.00--43.86 & 21 & 72.5 & 33.3--95.2 & 95.2 & 81.0 \\
  43.88--52.27 & 21 & 74.6 & 33.3--95.2 & 95.2 & 95.2 \\
  52.67--108.00 & 21 & 48.1 & 19.0--81.0 & 81.0 & 81.0 \\
  \bottomrule
  \end{tabular}
\end{table}

\begin{table}[H]
  \centering
  \caption{\textbf{Literature metal-containing structures by coordination number.} The 156 structures are grouped by the CN of the defining metal centre. CN counts coordination sites, with an $\eta$-bound group counting as one site, and is not the number of independent ligand molecules. Rates include verifier errors in each full subset denominator. Mean and range refer to the nine LLMs.}
  \label{tab:coordination-difficulty}
  \small
  \renewcommand{\arraystretch}{1.25}
  \begin{tabular}{@{}lrrrrr@{}}
  \toprule
  Group & $n$ & \makecell{Mean of\\nine (\%)} & Range (\%) & \makecell{gpt-6-astra\\(\%)} & \makecell{gpt-6.1-sol\\(\%)} \\
  \midrule
  2 & 15 & 50.4 & 33.3--86.7 & 80.0 & 86.7 \\
  3 & 22 & 56.6 & 18.2--95.5 & 95.5 & 81.8 \\
  4 & 53 & 56.4 & 13.2--83.0 & 83.0 & 83.0 \\
  5 & 35 & 35.6 & 8.6--62.9 & 62.9 & 51.4 \\
  6 & 21 & 52.4 & 19.0--81.0 & 81.0 & 76.2 \\
  7 & 6 & 59.3 & 33.3--100.0 & 100.0 & 83.3 \\
  8 & 4 & 61.1 & 0.0--100.0 & 100.0 & 100.0 \\
  \bottomrule
  \end{tabular}
\end{table}

\begin{table}[H]
  \centering
  \caption{\textbf{Resolve rate by total atom count, including hydrogens.} Bins are quintiles of atom count over all 300 reference structures. Rates include verifier errors in each full subset denominator. Mean and range refer to the nine LLMs.}
  \label{tab:atom-difficulty}
  \small
  \renewcommand{\arraystretch}{1.25}
  \begin{tabular}{@{}lrrrrr@{}}
  \toprule
  Group & $n$ & \makecell{Mean of\\nine (\%)} & Range (\%) & \makecell{gpt-6-astra\\(\%)} & \makecell{gpt-6.1-sol\\(\%)} \\
  \midrule
  18--51 & 59 & 68.7 & 39.0--94.9 & 94.9 & 89.8 \\
  52--66 & 58 & 67.4 & 34.5--87.9 & 87.9 & 84.5 \\
  67--80 & 64 & 63.5 & 26.6--85.9 & 85.9 & 82.8 \\
  81--107 & 57 & 52.6 & 19.3--80.7 & 80.7 & 77.2 \\
  108--202 & 62 & 42.8 & 12.9--75.8 & 75.8 & 69.4 \\
  \bottomrule
  \end{tabular}
\end{table}

\FloatBarrier   

\subsection{Two-tool ablation}
\label{app:ablation}

\begin{figure}[H]
    \centering
    \captionsetup{skip=4pt}
    \includegraphics[width=\textwidth]{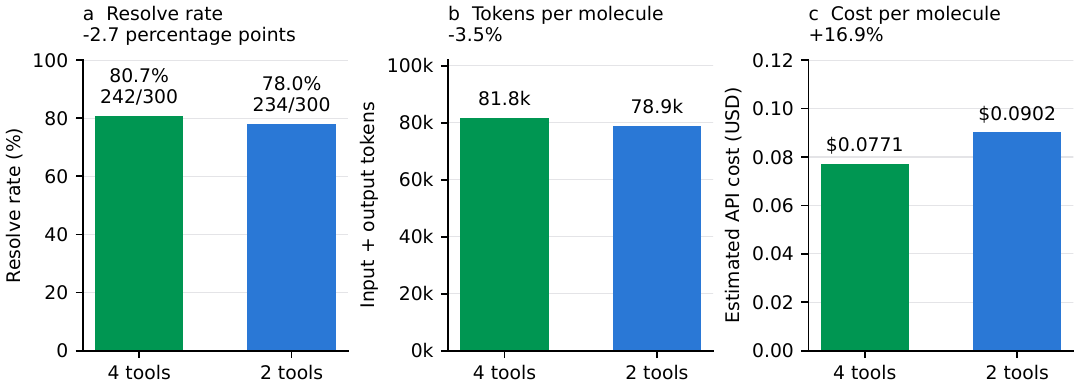}
    \caption{\textbf{GPT-6.1-sol with four versus two tools.} The two-tool agent exposes \texttt{view\_png\_file} and \texttt{bash\_kernel}; the four-tool configuration also exposes \texttt{view\_molecule} and \texttt{constraint\_xtb}. Panels compare resolve rate, total input plus output tokens per molecule, and estimated API cost per molecule. All denominators are 300; cached input is counted once. Changes are for two tools relative to four.}
    \label{fig:tool-ablation}
\end{figure}

\textbf{Fewer passes with two tools, despite lower token use.}
The four-tool configuration resolves 242 of 300 molecules (80.7\%), while the two-tool configuration resolves 234 (78.0\%), a difference of 2.7 percentage points (\Cref{fig:tool-ablation}).
Both configurations use the same molecular-identity, stereochemistry, and ligand-geometry criteria; every attempt remains in the denominator.
Mean input plus output tokens decrease from 81.8k to 78.9k per molecule ($-3.5\%$), while estimated API cost increases from \$0.0771 to \$0.0902 per molecule ($+16.9\%$).

\textbf{Per-molecule outcomes and token composition.}
The configurations jointly resolve 219 molecules; 23 are resolved only with four tools and 15 only with two, while neither resolves the remaining 43.
The two-tool run uses 14.0\% more uncached input tokens and 21.7\% more output tokens, despite its lower total token count.
This comparison covers one run per configuration with gpt-6.1-sol.

\FloatBarrier
\section{Additional qualitative examples}
\label{app:qualitative}
\begingroup
\small
\setlength{\parskip}{4pt}
Ten further examples, chosen to cover the failure categories of \Cref{tab:results} and the ways systems recovered; for each, the LLM, the molecule, and how many of the nine LLMs resolve it.

\textbf{A configuration invented for a perspective drawing: gpt-6-astra on an iridium metallacycle} (literature; 3 of 9).
The figure draws a metallacycle in perspective with bold front bonds.
The agent set the configuration from depth values it assigned to the atoms by hand (``Stereo specification from the perspective bonds in the drawing''); the submitted structure passes connectivity but has one inverted tetrahedral frame among the two checked.

\textbf{A misread label: gpt-6-sol on a rhodium Cp complex} (literature; 7 of 9).
The agent wrote the substituted Cp ring as a SMILES string with a third \ce{CF3} group where the drawing has a methyl beside a crowded \ce{F3C} label; the element composition was off by three fluorines and three hydrogens, and the structure failed before any alignment.

\textbf{A ligand that moved during relaxation: claude-opus-5-5 on a rhodium dibenzyl complex} (VSEPR-curated, square pyramidal; 6 of 9).
In five model turns the agent wrote one SMILES string and optimised it with GFN2-xTB without restraints.
One benzyl group slid towards $\eta^3$ coordination (Rh--C distances of 2.37, 2.37, and 2.27~\AA, against 2.08 to Rh--\ce{CH2} in the reference); the submitted structure fails the coordination-angle check. The agent noted that it had not set the arrangement at rhodium explicitly.

\textbf{The same molecule built two ways: a molybdenum alkoxide complex} (VSEPR-curated, square pyramidal; 3 of 9).
qwen3.8-max built the coordinates in NumPy without a SMILES string, and twice caught its own errors from the renders, ``a duplicated THF oxygen (I added O twice)'' and a ring that ``folded back onto the metal---my sp2 directions at C$\alpha$ had a sign error''; it was resolved.
claude-opus-5-5 wrote one SMILES string, optimised it without restraints, and saw that one Mo--O bond had stretched from 1.98 to 2.27~\AA{} (``quite long for an Mo--alkoxide bond''), but submitted it; the coordination angles differ by 28.0 degrees RMS and 63.4 degrees at maximum.

\textbf{Recovering an iridium Cp* complex after a failed embedding: claude-opus-5-5} (literature; 4 of 9).
After an embedding from SMILES, a distance check showed ``the Cp* ended up far from Ir, so the embedding is wrong''; the agent rebuilt the ligand around iridium, placed the Cp* ring by hand, and relaxed it under restraints.
The submitted structure passes connectivity, the checked stereochemistry and the reference-site ligand-geometry check.

\textbf{A charged open-shell dicobalt species: gpt-6-astra} (literature; $q = -1$, $S = 5/2$; 2 of 9).
The agent wrote the ligand as a SMILES string with the two cobalt ions unbonded, then placed them with a GFN2-xTB optimisation at the charge and spin printed on the depiction, under Co--N and Co--P distance restraints.
The submitted structure passes connectivity and stereochemistry but fails ligand geometry at one cobalt centre: the angular RMS deviation is $22.6^\circ$ (limit $20^\circ$), and the largest individual deviation is $48.3^\circ$ (limit $60^\circ$). Of the other models, two resolve the structure, four fail molecular identity and two return no structure.

\textbf{Out of budget: claude-opus-5 on the same species} (literature; 2 of 9).
The agent spent its 100 model calls reading the drawing---46 image views and some 48 crops, an attempt to install an image library, and a skeletonisation routine written by hand---and never wrote a structure; the session is scored as no answer.

\textbf{A perspective cage: gpt-6.1-sol on morphine} (natural product; 5 of 9).
The agent read the wedges and hashes literally off SynArchive's perspective drawing and built the exact enantiomer, all five centres inverted, although midway through the session it had typed a SMILES string identical to the reference and printed its CIP labels without comparing them with its build; gpt-6-sol built the same enantiomer.

\textbf{A self-check that repeats the error: claude-fable-5-1 on aplysistatin} (natural product; 8 of 9).
The agent read a hashed hydrogen on the lactone carbon as a wedge, and its own check re-encoded the same reading (``H on C8 wedge''), after which it concluded that ``the stereochemistry derived directly from the drawing's wedges matches my SMILES''; one centre of four is inverted.

\textbf{Two ligands read as one: kimi-k3 on a chromium(III) complex} (literature; $q = +3$, $S = 3/2$; 7 of 9).
The figure draws one of two tridentate ligands in perspective; the agent read the pair as a single macrobicycle with ``two such TBD caps (top/bottom)'', and the structure lacks twelve carbons, two nitrogens, and fourteen hydrogens, although the charge and spin it passed to xTB were right.
\endgroup


\end{document}